\documentstyle{ajour}

\begin{document}
\reversemarginpar

%


\authorrunninghead{Sriramkumar and Padmanabhan}
\titlerunninghead{Probes of the vacuum structure}




\title{Probes of the vacuum structure of quantum fields\\  
in classical backgrounds}

\author{L.~Sriramkumar}
\affil{Racah Institute of Physics, Hebrew University,\\ 
Givat Ram, Jerusalem 91904, Israel\thanks{Present 
Address:~Theoretical Physics Institute, Department of Physics, 
University of Alberta, Edmonton, Alberta~T6G~2J1, Canada. 
E-mail:~slakshm@phys.ualberta.ca.}.}

\and

\author{T.~Padmanabhan}
\affil{IUCAA, Post Bag 4, Ganeshkhind, Pune 411 007, India.}
\email{paddy@iucaa.ernet.in.}

\abstract{We compare the different approaches 
presently available in literature to probe the vacuum structure 
of quantum fields in classical electromagnetic and gravitational 
backgrounds.
We compare the results from the Bogolubov transformations and 
the effective Lagrangian approach with the response of monopole 
detectors (of the Unruh-DeWitt type) in non-inertial frames in 
flat spacetime and in inertial frames in different types of 
classical electromagnetic backgrounds.
We also carry out such a comparison in inertial and rotating
frames when boundaries are present in flat spacetime.
We find that the results from these different approaches do 
not, in general, agree with each other.
We attempt to identify the origin of these differences and then 
go on to discuss its implications for classical gravitational 
backgrounds.}


\begin{article}
\newpage

\section{Introduction}\label{sec:intro}

The vacuum state of a quantum field develops a non-trivial 
structure in a classical electromagnetic or gravitational 
background.
As a result, essentially, two different types of phenomena 
occur in a classical background: (i)~polarization of the 
vacuum and (ii)~production of particles corresponding to 
the quantum field.
Apart from these two effects, there is another feature that 
one encounters in a gravitational background:~the concept
of a particle turns out to be coordinate dependent.
(For a detailed discussion on these different aspects of 
quantum field theory in classical backgrounds, see the following 
books~\cite{bd82,gmr85,fulling89,fgs91,wald94,gmm94,mt97} 
and the recent reviews~\cite{bmps95,ford97}.)
A classic example of vacuum polarization is the Casimir
effect~\cite{casimir48} while Hawking radiation from collapsing 
black holes is the most famous example of particle 
production~\cite{hawking75}.
The coordinate dependence of the particle concept that 
arises in a gravitational background is well illustrated
by the fact that the Rindler vacuum turns out to be
inequivalent to the Minkowski vacuum~\cite{fulling73}.

Different approaches have been formulated in literature 
to study the evolution of a quantum field in a classical 
electromagnetic or gravitational background. 
On the one hand, the Bogolubov transformations~\cite{bogolubov58}
and the effective Lagrangian 
approach~\cite{he36,schwinger51,dewitt75} offer us formal 
methods to probe the vacuum structure of the quantum field.
On the other, studying the response of detectors coupled to 
the quantum field provides us with an operational tool for 
understanding the concept of a particle~\cite{unruh76,dewitt79}.
Often in literature, one of these approaches has been used to 
study the behavior of a quantum field in a classical background 
and, apart from a few instances (see, for e.g.,  
Refs.~\cite{letaw81,paddy82,letpfa81,go83,grove86}), the results 
from these different approaches have not been compared.
Due to this reason, the possibility that these approaches can 
lead to different results has not been adequately emphasized.
(As we shall see later, these different approaches do, in 
general, lead to different results.)
Our motivation in this paper is to compare the results from 
these different approaches in a variety of situations, identify 
the origin of the differences that arise and understand its
implications for classical gravitational backgrounds. 

A detailed outline of the contents of this paper is as follows. 
In Section~\ref{sec:prbs}, we shall briefly review the three
different approaches that are available at present to study
the evolution of a quantum field in a classical background,
viz.~(i)~the Bogolubov transformations, (ii)~the response 
of detectors and (iii)~the effective Lagrangian approach. 
We shall confine our attention in this paper to monopole 
detectors of the Unruh-DeWitt type.

In Section~\ref{sec:nifrms}, we shall compare the results 
from these different approaches in non-inertial frames in 
flat spacetime.
In Section~\ref{subsec:nitraj}, following 
Refs. \cite{letaw81,paddy82}, we construct 
different non-inertial trajectories in flat 
spacetime which are integral curves of 
timelike Killing vector fields. 
In Section~\ref{subsec:nifrmscomp}, we evaluate the 
effective Lagrangian in coordinates adapted to these 
trajectories.
We compare these results with the results from the 
Bogolubov transformations and the response of the 
Unruh-DeWitt detector that have been obtained in 
literature before~\cite{letaw81,paddy82,letpfa81}.
In Section~\ref{subsec:bogdet}, we express the response 
of the Unruh-DeWitt detector in terms of the Bogolubov 
coefficients~\cite{go83} and identify the origin of the 
differences that arise in the results from these two 
approaches.

In Section~\ref{sec:bndrs}, we shall carry out such a 
comparison when boundaries are present in flat spacetime.
In Section~\ref{subsec:casine}, we compare the response 
of an inertial Unruh-DeWitt detector in the Casimir vacuum
with the result obtained from the effective Lagrangian approach. 
In Section~\ref{subsec:casrot}, we briefly discuss as 
to how the response of a rotating detector would compare 
with the effective Lagrangian when a boundary condition is 
imposed on the horizon in the rotating frame. 

In Section~\ref{sec:embckgnds}, we shall compare the 
results in inertial frames in different types of 
classical electromagnetic backgrounds.
In Section~\ref{subsec:nlcd}, following Ref.~\cite{sriram99},
we discuss the response of a monopole detector that is coupled 
to the quantum field through a gauge-invariant and non-linear 
interaction.
In Sections~\ref{subsec:tdef}, \ref{subsec:tindef} 
and~\ref{subsec:tindmf}, we study the response of 
this detector (when it is in inertial motion) in 
a time-dependent electric field, a time-independent 
electric field and a time-independent magnetic field,
backgrounds, respectively.
We also discuss as to how the response of the detector 
compares with the results expected from the Bogolubov 
transformations and the effective Lagrangian approach.

In the concluding Section~\ref{sec:dscssn}, we shall 
first briefly summarize the results of our analysis 
in Section~\ref{subsec:what}.
Then, in Section~\ref{subsec:implctns}, we shall go 
on to discuss the implications of these results for 
classical gravitational backgrounds.

Our conventions and notations are as follows.
Throughout this paper, we shall set $\hbar=c=1$. 
We shall always work in $(3+1)$~dimensions and the metric 
signature we shall adopt is $(+, -, -, -)$.
Also, for the sake of convenience and clarity in notation, 
we shall denote the set of coordinates~$x^{\mu}$ 
as~${\tilde x}$ and we shall write the derivative 
$(\partial/\partial x)$ simply as $\partial_{x}$.
Finally, we shall denote complex conjugation and Hermitian 
conjugation by an asterisk and a dagger, respectively.

\section{Probes of the vacuum structure}\label{sec:prbs}

In this section, we shall briefly review the three different
probes of the vacuum structure of quantum fields in classical
backgrounds, viz.~(i)~the structure of the Bogolubov 
transformations, (ii)~the response of the Unruh-DeWitt detector 
and (iii)~the effective Lagrangian approach.
We shall gather here the results that will prove to be 
essential for our discussion later on.

\subsection{Bogolubov transformations}\label{subsec:bgbv}

Consider a quantum scalar field ${\hat \Phi}$ of mass $m$ 
evolving in a given classical background.
Let the quantum field ${\hat \Phi}$ satisfy the following 
equation of motion:
\begin{equation}
\left({\hat H}+m^2\right){\hat \Phi}=0,\label{eqn:mtn}
\end{equation}
where ${\hat H}$ is a differential operator whose form 
depends on the classical background.
A conserved current corresponding to this equation of motion
can then be used to define a scalar product for the modes of 
the quantum field.
Let $\left\{u_{i}({\tilde x})\right\}$ and 
$\left\{{\bar u}_{k}({\tilde x})\right\}$ 
be two complete sets of positive {\it norm},\/ orthonormal  
modes corresponding to such a scalar product\footnote{The 
definition of positive norm modes will {\it not},\/ in general, 
coincide with the definition of positive frequency modes. It is 
the former property rather than the latter that has to be taken 
into account in constructing the Fock space of a quantum field.}.
When two such complete sets of modes exist, one set of
modes can be expressed in terms of the other using the 
Bogolubov transformations as follows (see, for e.g., 
Ref.~\cite{bd82}, Sec.~3.2):
\begin{equation}
{\bar u}_k({\tilde x})
= \sum_{i} \left[\alpha_{ki}\, u_i({\tilde x}) 
+ \beta_{ki}\, u_i^*({\tilde x})\right],\label{eqn:bglbvtrnsfmtn1} 
\end{equation}
or conversely
\begin{equation}
u_i({\tilde x})
= \sum_{k} \left[\alpha_{ki}^*\, {\bar u}_k({\tilde x}) 
- \beta_{ki}\, {\bar u}_k^*({\tilde x})\right].
\label{eqn:bglbvtrnsfmtn2}
\end{equation}
The quantities $\alpha_{ki}$ and $\beta_{ki}$ are 
called the Bogolubov coefficients~\cite{bogolubov58}.
Using the orthonormality of the modes and the 
relation~(\ref{eqn:bglbvtrnsfmtn1}), the 
Bogolubov coefficients can be expressed as 
\begin{equation}
\alpha_{ki} = \left({\bar u}_k({\tilde x}), 
u_i({\tilde x})\right) 
\qquad{\rm and}\qquad
\beta_{ki} = -\left({\bar u}_k({\tilde x}), 
u_i^*({\tilde x})\right),
\label{eqn:bglbvcffcnts}
\end{equation}
where the brackets denote scalar products.

A {\it real}\/ quantum scalar field~${\hat \Phi}$, 
for instance, can be decomposed in terms of the 
two sets of modes $\left\{u_{i}({\tilde x})\right\}$ 
and $\left\{{\bar u}_{k}({\tilde x})\right\}$ as 
follows:
\begin{equation}
{\hat \Phi}({\tilde x})
=\sum_{i}\left[{\hat a}_{i}\, u_{i}({\tilde x})
+{\hat a}_{i}^{\dag}\, u_{i}^{*}({\tilde x})\right]
\end{equation}
and
\begin{equation}
{\hat \Phi}({\tilde x})
=\sum_{k}\left[{\hat {\bar a}}_{k}\, 
{\bar u}_{k}({\tilde x})
+{\hat {\bar a}}_{k}^{\dag}\, 
{\bar u}_{k}^{*}({\tilde x})\right].
\end{equation}
Using these expansions and the Bogolubov 
transformations~(\ref{eqn:bglbvtrnsfmtn2}), 
it can be easily shown that
\begin{equation}
{\hat {\bar a}}_k 
= \sum_{i}\, \left(\alpha_{ki}^*\, {\hat a}_i
- \beta_{ki}^*\, {\hat a}_i^{\dagger}\right).
\end{equation}
It is clear from this expression that the Fock spaces based 
on the two sets of modes~$\left\{u_{i}({\tilde x})\right\}$ 
and~$\left\{{\bar u}_{k}({\tilde x})\right\}$ will prove to 
be different whenever the Bogolubov coefficient~$\beta$ 
turns out to be non-zero.
When~$\beta$ is non-zero, the expectation 
value of the number operator $\left({\hat {\bar a}}^{\dag}_{k}
{\hat {\bar a}}_{k}\right)$ in the vacuum state annihilated by 
the operator~${\hat a}_i$ is given by
\begin{equation}
\left\langle\left({\hat {\bar a}}^{\dag}_{k}
{\hat {\bar a}}_{k}\right)\right\rangle
=\sum_{i} \vert \beta_{ki}\vert^2.\label{eqn:expval} 
\end{equation}

In a gravitational background, the Bogolubov transformations
can either relate the modes of a quantum field at two different 
times in the same coordinate system or the modes in two different 
coordinate systems covering the same region of spacetime.
When the Bogolubov coefficient~$\beta$ is non-zero, in the 
latter context, such a result is normally interpreted as 
implying that the quantization in the two coordinate 
systems are inequivalent~\cite{fulling73}.
Whereas, in the former context, a non-zero~$\beta$ is
attributed to the production of particles by the background 
gravitational field~\cite{parker77}.
Similarly, in an electromagnetic background, a 
non-zero~$\beta$ relating the modes of a quantum field at 
different times (in a particular gauge) implies that the 
background produces particles (see, for instance, 
Ref.~\cite{fgs91}, Sec.~2.1).
Though it has been suggested in literature that inequivalent
(i.e. gauge-dependent) vacua may arise in  electromagnetic 
backgrounds as well, it has not been explicitly shown as yet 
(see Ref.~\cite{paddy90}; also see Ref.~\cite{paddy91}, 
Sec.~4.6).

\subsection{Response of the Unruh-DeWitt 
detector}\label{subsec:udd}

A detector is an idealized point like object whose motion 
is described by a classical worldline, but which nevertheless
possesses internal energy levels.
Such detectors are essentially described by the interaction
Lagrangian for the coupling between the degrees of freedom
of the detector and the quantum field.
The simplest of the different possible detectors is the 
detector due to Unruh and DeWitt~\cite{unruh76,dewitt79}. 
Consider a Unruh-DeWitt detector that is moving along a 
trajectory ${\tilde x}(\tau)$, where $\tau$ is the proper 
time in the frame of the detector.
The interaction of the Unruh-DeWitt detector with a 
{\it real}\/ scalar field~$\Phi$ is described by the 
interaction Lagrangian
\begin{equation}
{\cal L}_{\rm int}
= {\bar c}\, \mu(\tau)\, \Phi\left[{\tilde x}(\tau)\right],
\label{eqn:lint}
\end{equation}
where ${\bar c}$ is a small coupling constant and $\mu$ 
is the detector's monopole moment.
Let us assume that the quantum field ${\hat \Phi}$ is 
initially in the vacuum state $\vert 0\rangle$ and the 
detector is in its ground state~$\vert {\bar E}_0\rangle$ 
corresponding to an energy eigen value~${\bar E}_0$.
Then, up to the first order in perturbation theory, the 
amplitude of transition of the Unruh-DeWitt detector to 
an excited state~$\vert {\bar E}\rangle$,  corresponding 
to an energy eigen value~${\bar E}\, (>{\bar E}_0)$, is 
described by the integral (see, for instance, 
Ref.~\cite{bd82}, Sec.~3.3)
\begin{equation}
{\cal A}({\cal E}) = {\cal M}
\int\limits_{-\infty}^{\infty} d\tau\, e^{i {\cal E}\tau}\, 
\langle\Psi\vert{\hat \Phi}[{\tilde x}(\tau)]\vert 0\rangle,
\label{eqn:detamp}
\end{equation}
where ${\cal M}\equiv i{\bar c}\, \langle {\bar E}\vert 
{\hat \mu}(0) \vert {\bar E}_{0}\rangle$,
${\cal E}=({\bar E}-{\bar E}_0)>0$ and $\vert \Psi\rangle$ 
is the state of the quantum scalar field after its interaction 
with the detector.
(Since the quantity ${\cal M}$ depends only on the internal 
structure of the detector and does not depend on its motion, 
we shall drop this quantity hereafter.) 
The transition probability of the detector to all possible 
final states $\vert \Psi\rangle$ of the quantum field is 
given by
\begin{equation}
{\cal P}({\cal E}) 
=\sum_{\vert\Psi\rangle}{\vert {\cal A}({\cal E})\vert}^2
= \int\limits_{-\infty}^\infty d\tau\, 
\int\limits_{-\infty}^\infty d\tau'\, 
e^{-i{\cal E}(\tau-\tau')}\, 
G^{+}\left[{\tilde x}(\tau), {\tilde x}(\tau')\right],
\label{eqn:detprob}
\end{equation}
where $G^{+}\left[{\tilde x}(\tau), {\tilde x}(\tau')\right]$ 
is the Wightman function defined as
\begin{equation}
G^{+}\left[{\tilde x}(\tau), {\tilde x}(\tau')\right]
=\langle 0 \vert 
{\hat \Phi}\left[{\tilde x}(\tau)\right]\,
{\hat \Phi}\left[{\tilde x}(\tau')\right]
\vert 0 \rangle.\label{eqn:wgfn}
\end{equation}
For trajectories which are integral curves of timelike 
Killing vector fields, the Wightman function will be 
invariant under time translations in frame of the detector.
In such a case, a transition probability rate for the detector
can be defined as follows:
\begin{equation}
{\cal R}({\cal E}) 
= \int\limits_{-\infty}^\infty d\Delta\tau\,
e^{-i{\cal E}\Delta\tau}\, G^{+}(\Delta\tau),
\label{eqn:detrate}
\end{equation} 
where $\Delta\tau = (\tau - \tau')$.

\newpage
\subsection{The effective Lagrangian 
approach}\label{subsec:efflag}

The effective Lagrangian approach consists of integrating 
out the degrees of freedom corresponding to the quantum 
field thereby obtaining a correction to the Lagrangian 
describing the classical background~\cite{he36}. 
The correction thus obtained, in general, has a real as well 
as an imaginary part to it~\cite{schwinger51,dewitt75}. 
Its real part is interpreted as the `vacuum-to-vacuum' 
transition amplitude, i.e. the amplitude for the quantum 
field to remain in the initial vacuum state at late 
times and the existence of a non-zero imaginary part 
is attributed to the instability of the vacuum.
In other words, the real part part of the effective 
Lagrangian reflects the amount of vacuum polarization 
and the imaginary part is related to the number of 
particles produced by the classical background. 

Consider the case of a {\it real}\/ quantum scalar 
field~${\hat \Phi}$ satisfying the equation of 
motion~(\ref{eqn:mtn}) in a given classical background. 
For such a case, the correction to the Lagrangian 
describing the classical background is obtained by 
integrating the degrees of freedom corresponding to 
the quantum field~${\hat \Phi}$.
In Schwinger's proper time formalism, the correction is 
given by the integral~\cite{schwinger51,dewitt75}
\begin{equation}
{\cal L}_{\rm corr}=-\left(\frac{i}{2}\right)
\int\limits_0^{\infty}\frac{ds}{s}\;e^{-im^2s}\;
K({\tilde x}, {\tilde x}; s),\label{eqn:lcorr}
\end{equation}
where $K({\tilde x}, {\tilde x}; s)$ is the 
${\tilde x'}\to {\tilde x}$ limit of the quantity
\begin{equation}
K({\tilde x}, {\tilde x'}; s) 
\equiv \langle {\tilde x}\vert 
e^{-i{\hat H}s} \vert {\tilde x'}\rangle.
\label{eqn:kernel}
\end{equation}
The quantity $K({\tilde x}, {\tilde x'}; s)$ is the path 
integral kernel of a quantum mechanical system described 
by the time evolution operator~${\hat H}$ and the integration 
variable $s$ acts as the time parameter for the quantum 
mechanical system.
The integral~(\ref{eqn:lcorr}) yields a divergent expression 
even in the Minkowski vacuum in flat spacetime.
Therefore, the effective Lagrangian for any non-trivial 
background has to be regularized by subtracting this 
contribution due to flat spacetime.

Schwinger's proper time formalism can also be used to 
evaluate the Feynman propagator.
The Feynman propagator corresponding to a 
quantum field~${\hat \Phi}$ satisfying the equation 
of motion~(\ref{eqn:mtn}) is described by the 
following integral~\cite{schwinger51,dewitt75}:
\begin{equation}
G_{\rm F}({\tilde x},{\tilde x'})
=-i\int\limits_0^{\infty}ds\;
e^{-i\left[m^2s-i(\epsilon/s)\right]}\;
K({\tilde x}, {\tilde x'}; s),\label{eqn:fgfn}
\end{equation}
where $\epsilon \to 0^{+}$ and $K({\tilde x},
{\tilde x'};s)$ is the quantum mechanical kernel 
defined in Eq.~(\ref{eqn:kernel}).

\section{In non-inertial frames in flat 
spacetime}\label{sec:nifrms}

Earlier, in Section~\ref{subsec:udd}, we had mentioned 
that if the trajectory of the Unruh-DeWitt detector is 
chosen to be an integral curve of a timelike Killing 
vector field, then the Wightman function will be invariant 
under translations in the proper time in the frame of 
the detector.
We had also pointed out that in such a case we can define 
a transition probability rate for the detector.
In Section~\ref{subsec:nitraj}, we shall construct integral 
curves of timelike Killing vector fields in flat spacetime 
and, as we shall see, these curves correspond to different 
types of non-inertial trajectories.
Then, in Section~\ref{subsec:nifrmscomp}, we shall go on 
to compare the response of Unruh-DeWitt detectors with the 
results from the Bogolubov transformations and the effective 
Lagrangian approach in coordinate systems adapted to these
non-inertial trajectories.

\subsection{Stationary trajectories in flat 
spacetime}\label{subsec:nitraj}

As is well known, there are ten independent timelike Killing 
vector fields in flat spacetime.
They correspond to three types of symmetries---translations, 
rotations and boosts.
Different types of trajectories can be generated by choosing
various linear combinations of these Killing vector fields.
However, we do not gain anything by treating, say, boosts 
along the three different axes separately.
A sufficiently general Killing vector field in flat spacetime 
that incorporates effects of translations, rotations and boosts
can be written as~\cite{letaw81,paddy82}
\begin{equation}
\xi^{\mu}({\tilde x}) 
=\left(1+\kappa x, \kappa t-\lambda y, 
\lambda x-\rho z,\rho y\right),
\label{eqn:gnrlkvf}
\end{equation}
where $\kappa$, $\lambda$ and $\rho$ are constants and 
$(t,x,y,z)$ are the Minkowski coordinates.

Let us now consider some special cases 
of~$\xi^{\mu}({\tilde x})$ and the 
trajectories generated by them.
The simplest of the cases is when $\kappa$, $\lambda$ and 
$\rho$ are all set to zero.
For such a case, the Killing vector 
field~$\xi^{\mu}({\tilde x})$ reduces to 
\begin{equation}
\xi^{\mu}({\tilde x})=(1,0,0,0).
\end{equation}
The natural coordinate systems corresponding to this
Killing  vector field are the rectangular Minkowski 
coordinates and the other curvilinear coordinates.
The flat space line element in terms of the Minkowski 
coordinates is given by
\begin{equation}
ds^2=dt^2-d{\bf x}^2,\label{eqn:minkmet}
\end{equation}
where ${\bf x}\equiv(x,y,z)$.
Other than the inertial trajectory we have just discussed, 
the Killing vector field~$\xi^{\mu}({\tilde x})$ also 
generates five different types of non-inertial 
trajectories~\cite{letaw81,paddy82}.
We shall consider three of them here.

\subsubsection{Uniformly accelerated 
motion}\label{subsubsec:rind}

Let us choose $\lambda=\rho=0$.
For such a case, the Killing vector 
field~$\xi^{\mu}({\tilde x})$ reduces to
\begin{equation}
\xi^{\mu}({\tilde x})
=\left(1+\kappa x,\kappa t,0,0\right).
\end{equation}
The integral curve of such a Killing vector field is 
given by
\begin{equation}
{\tilde x}(\tau)
=\kappa^{-1}\biggl({\rm sinh}(\kappa\tau), 
{\rm cosh}(\kappa\tau),0,0\biggl)
\label{eqn:uniacctraj}
\end{equation}
which corresponds to the trajectory of a uniformly 
accelerated observer moving with a proper 
acceleration~$\kappa$.
A natural coordinate system for such an observer is 
related to the Minkowski coordinates by the following
transformations:
\begin{equation}
t= g^{-1}\, \xi\, \sinh(g\eta)\;\; ;\;\;\; 
x= g^{-1}\, \xi\, \cosh(g\eta)\;\; 
;\;\; y=y \;\; ; \;\; z=z,\label{eqn:rindtransf} 
\end{equation}
where $g$ is a constant.
The new coordinates $(\eta, \xi, y, z)$ are called the 
Rindler coordinates~\cite{rindler66} and the proper 
acceleration of an observer at the point~$\xi$ in this 
coordinate system is~$\left(g/\xi\right)$.
In terms of the Rindler coordinates, the flat spacetime 
line element~(\ref{eqn:minkmet}) is given by
\begin{equation}
ds^2= \xi^2\, d\eta^2 
- g^{-2}\, d\xi^2 - dy^2 - dz^2.\label{eqn:rindmet}
\end{equation}

\subsubsection{Rotational motion}\label{subsubsec:rot} 

On setting $\rho=0$ in Eq.~(\ref{eqn:gnrlkvf}), we 
obtain that
\begin{equation}
\xi^{\mu}({\tilde x})
=\left(1+\kappa x,\kappa t- \lambda y,\lambda x, 0\right).
\end{equation}
The trajectory generated by such a Killing vector field is 
given by
\begin{equation}
{\tilde x}(\tau)
=\sigma^{-2}\biggl(\lambda \sigma\tau, 
\kappa \cos(\sigma\tau), 
\kappa \sin(\sigma\tau),0\biggl),
\label{eqn:rottraj}
\end{equation}
where $\sigma^{2}=(\lambda^2-\kappa^2)$ and 
$\vert\kappa\vert < \vert\lambda\vert$.
This trajectory corresponds to that of an observer moving 
with a linear velocity $(\kappa/\lambda)$ along a circle 
of radius $(\kappa/\sigma^2)$.
The coordinates~$(t,r,\theta,z)$ of an observer rotating 
about the $z$-axis with an angular frequency~$\Omega$ are 
related to the Minkowski coordinates by the following 
transformations:
\begin{equation}
t=t\;\;;\;\; x=r\cos(\theta + \Omega t)\;\;;\;\;
y=r\sin(\theta+ \Omega t)\;\;;\;\;z=z.
\label{eqn:rottransf}
\end{equation}
In the rotating coordinate system, flat spacetime is 
described by the line element 
\begin{equation}
ds^2=dt^2-dr^2-r^2\,\left(d\theta+\Omega\, dt\right)^2-dz^2.
\label{eqn:rotmet}
\end{equation}

\subsubsection{A cusped motion}\label{subsubsec:cusp}

On setting  $\lambda=\kappa$ and $\rho=0$, the Killing
vector field~$\xi^{\mu}({\tilde x})$ reduces to
\begin{equation}
\xi^{\mu}({\tilde x})
=\left(1+\kappa x, \kappa t-\kappa y,\kappa x,0\right).
\end{equation}
This Killing vector field gives rise to a peculiar cusped 
motion with the trajectory 
\begin{equation}
{\tilde x}(\tau)
=\left(\tau+(\kappa^2\tau^3/6), (\kappa\tau^2/2), 
(\kappa^2\tau^3/6),0\right).\label{eqn:cusptraj}
\end{equation}
A natural coordinate system corresponding to an observer in
motion along such a trajectory is related to the Minkowski 
coordinates by the following transformations:
\begin{eqnarray}
t= \left(a^2{\bar t}^3/6\right)
+\left[a{\bar x}+(1/2)\right]\, {\bar t}+{\bar y}\;\;&;&\;\;
x=\left(a {\bar t}^2/2\right)
+\left[{\bar x}-(a/2)\right]\nonumber\\
y=\left(a^2{\bar t}^3/6\right)
+\left[a{\bar x}-(1/2)\right]\, {\bar t}
+{\bar y}\;\;&;&\;\; z=z,\label{eqn:cusptransf}
\end{eqnarray}
where~$a$ is a constant.
The flat spacetime line element in terms of the new 
coordinates $({\bar t}, {\bar x},{\bar y}, z)$ is given by
\begin{equation}
ds^2=2a\,{\bar x}\,d{\bar t}^2+2\,d{\bar y}\,d{\bar t} 
-d{\bar x}^2-dz^2.\label{eqn:cuspmet}
\end{equation}
For want of a better name, we shall hereafter refer to 
the coordinates $({\bar t}, {\bar x},{\bar y}, z)$ as 
the `cusped' coordinates. 

\subsection{Comparison}\label{subsec:nifrmscomp}

The quantum field we shall consider in this section is a 
{\it real}\/ and massless scalar field~$\Phi$ described 
by the action
\begin{equation}
{\cal S}[\Phi]=\left(\frac{1}{2}\right)
\int d^4x\; \sqrt{-g}\; 
\biggl(g_{\mu\nu}\,\partial^{\mu}\Phi\,
\partial^{\nu}\Phi\biggl),
\label{eqn:gravact}
\end{equation}
where $g_{\mu\nu}$ is the metric tensor describing the 
classical gravitational background. 
Varying this action leads to an equation of motion such 
as~(\ref{eqn:mtn}) with $m$ set to zero and the 
operator~${\hat H}$ given by
\begin{equation}
{\hat H}\equiv\frac{1}{\sqrt{-g}}\,\partial_{\mu}
\left(\sqrt{-g}\,g^{\mu\nu}\partial_{\nu}\right).
\label{eqn:gravop}
\end{equation}
Let us now assume that the massless quantum scalar 
field~${\hat \Phi}$ is in the Minkowski vacuum state. 
For such a case, the Wightman function~(\ref{eqn:wgfn}) 
in terms of the Minkowski coordinates is given by the 
following expression (see, for e.g., Ref.~\cite{bd82}, 
Sec.~3.3):
\begin{equation}
G^{+}({\tilde x},{\tilde x'})
=\left(\frac{-1}{4\pi^2}\right)
\left(\frac{1}{(t-t'-i\epsilon)^2
-\vert {\bf x}-{\bf x}'\vert^2}\right),
\label{eqn:wgfnine}
\end{equation}
where, as we had mentioned earlier, $\epsilon\to 0^{+}$.
The transition probability rate of the Unruh-DeWitt detector 
in the Minkowski vacuum when it is in motion along the  
non-inertial trajectories we had discussed in the last 
subsection is then obtained by substituting these trajectories 
in the above Wightman function and evaluating the 
integral~(\ref{eqn:detrate}).
These transition probability rates have already been 
evaluated in literature~\cite{unruh76,dewitt79,letaw81,paddy82}. 
The Bogolubov coefficients relating the modes in these 
non-inertial coordinate systems and the Minkowski modes 
have been obtained in literature as 
well~\cite{fulling73,paddy82,letpfa81}.
 
In what follows, we shall first evaluate the quantum 
mechanical kernel~$K({\tilde x}, {\tilde x'};s)$ (as
defined in Eq.~(\ref{eqn:kernel})) corresponding to 
the operator~${\hat H}$ (given by Eq.~(\ref{eqn:gravop}) 
above) in the non-inertial coordinate systems.
Substituting this kernel in Eq.~(\ref{eqn:fgfn}) we 
shall obtain the resulting Feynman propagator.
(Since evaluating the kernel and the corresponding Feynman 
propagator involves lengthy algebra we shall relegate 
the details of the calculation to the Appendix.)
Then, from the coincidence limit (i.e. when ${\tilde x'} 
= {\tilde x}$) of the kernel, we shall evaluate the 
effective Lagrangian using the expression~(\ref{eqn:lcorr}) 
and compare these results with the response of Unruh-DeWitt 
detectors and the results from the Bogolubov transformations.
We calculate the Feynman propagator using Schwinger's proper 
time formalism so that it can be compared with the Wightman 
function~(\ref{eqn:wgfnine}) evaluated along the trajectory 
of the detector.
[The boundary condition and the resulting pole structure
of the Wightman function is, of course, different from that 
of the Feynman propagator. 
In general, the correct boundary condition can always be 
identified by comparing the pole structure in the limit
of free field theory.
In this limit, the Wightman function should have the
term $(t-t'-i\epsilon)^2$ (cf.~Eq.~(\ref{eqn:wgfnine})), 
whereas the Feynman propagator will contain the term
$\left[(t-t')^2-i\epsilon\right]$ 
(cf.~Eq.~(\ref{eqn:fgfnine})).]
This check is to ensure that we are evaluating the effective
Lagrangian corresponding to the same conditions under which
the response of the Unruh-DeWitt detectors have been studied
in literature.

Before we go on to discuss the case of the non-inertial
trajectories, let us very briefly discuss the inertial case.
(The arguments we shall present here will prove to be useful 
for our discussion later on.)
Consider an inertial detector stationed at a point, 
say,~${\bf a}$. 
Let us now evaluate the transition amplitude (in fact, its 
complex conjugate) of this detector in the Minkowski vacuum.
It is easy to see from Eq.~(\ref{eqn:detamp}) that it is 
{\it only}\/ the positive norm modes of the quantum field
that contribute to the resulting integral. 
Therefore, the transition amplitude of the detector 
corresponding to a single mode~${\bf k}$ of the field 
is given by
\begin{eqnarray}
{\cal A}_{\bf k}^{*}({\cal E})
&=&\left(\frac{e^{i{\bf k}\cdot{\bf a}}}
{\sqrt{(2\pi)^3\, 2\omega_k}}\right)\,
\int\limits_{-\infty}^{\infty}dt\, 
\exp -\left[i(\omega_{k}+{\cal E})t\right]\nonumber\\
&=&\left(\frac{e^{i{\bf k}\cdot{\bf a}}}
{\sqrt{4\pi \omega_{k}}}\right)\,
\delta^{(1)}({\cal E}+\omega_{k}),\label{eqn:detampine}
\end{eqnarray}
where $\omega_k=\vert{\bf k}\vert$.
In the Minkowski coordinates, the definition of positive norm 
modes match the definition of positive frequency modes. 
Therefore, the quantity $\omega_{k}$ appearing in the delta 
function above is always greater than (or equal to) zero.  
Since ${\cal E}$ is greater than zero as well, the argument 
of the delta function is a positive definite quantity and 
hence the transition amplitude~${\cal A}_{\bf k}^{*}({\cal E})$ 
above reduces to zero for {\it all}\/ modes~${\bf k}$.
In other words, an inertial detector does not respond in the
Minkowski vacuum state.

The kernel~(\ref{eqn:kernel}) corresponding to the 
operator~${\hat H}$ (as defined in Eq.~(\ref{eqn:gravop})) 
in the Minkowski coordinates can be easily evaluated.
In the coincidence limit, this kernel reduces to
(cf.~Eq.~(\ref{eqn:minkkrnl}))
\begin{equation}
K({\tilde x},{\tilde x};s)
=\left(\frac{1}{16\pi^2 i s^2}\right)
\label{eqn:krnl0}
\end{equation} 
and the corresponding effective Lagrangian is given by
\begin{equation}
{\cal L}_{\rm corr}^0=-\left(\frac{1}{16\pi^2}\right)
\int\limits_{0}^{\infty}\frac{ds}{s^3}.\label{eqn:lcorr0}
\end{equation}
This quantity diverges near $s=0$ and, as we had pointed out 
in Section~\ref{subsec:efflag}, all other effective Lagrangians 
have to be regularized by subtracting this divergent expression.

\subsubsection{In the Rindler coordinates}

The Wightman function in the frame of a uniformly 
accelerated observer is obtained by substituting 
the trajectory~(\ref{eqn:uniacctraj}) in  
Eq.~(\ref{eqn:wgfnine}).
It is given by 
\begin{equation}
G^{+}({\tilde x},{\tilde x'})
=\left(\frac{-1}{4\pi^2}\right)
\sum_{n=-\infty}^{\infty}\left(\tau-\tau'-i\epsilon
+2\pi in \kappa^{-1}\right)^{-2}.
\end{equation}
(This Wightman function corresponds to the case 
wherein the quantities~$\eta$ and~$g$ in the Feynman 
propagator~(\ref{eqn:fgfnrind}) are set to~$(\tau/\xi)$ 
and~$(\kappa\xi)$, respectively.)
The resulting transition probability rate can be
evaluated to be~\cite{unruh76,dewitt79}
\begin{equation}
{\cal R}({\cal E})=\left(\frac{1}{2\pi}\right)
\left(\frac{{\cal E}}{e^{2\pi{\cal E}\kappa^{-1}}-1}\right),
\label{eqn:thrml}
\end{equation}
which is a thermal spectrum corresponding to a 
temperature~$T=(\kappa/2\pi)$.
The Bogolubov coefficient~$\beta$ relating the Rindler modes 
and the Minkowski modes turns out to be non-zero and, in fact,
the expectation value of the Rindler number operator in the 
Minkowski vacuum yields the above thermal spectrum as 
well~\cite{fulling73}.
However, on evaluating the kernel~(\ref{eqn:kernel}) in the 
Rindler coordinates, we find that, in the coincidence limit, 
it reduces to the kernel~(\ref{eqn:krnl0}) in the Minkowski 
coordinates (cf.~Eq.~(\ref{eqn:rindkrnl})). 
Therefore, the effective Lagrangian in the Rindler 
coordinates vanishes on regularization.

\subsubsection{In the rotating coordinates}\label{subsubsec:rotcomp}

On substituting the trajectory~(\ref{eqn:rottraj}) in 
Eq.~(\ref{eqn:wgfnine}), we find that the Wightman function
along the trajectory of a rotating detector is given by
\begin{equation}
\!\!\!\!\!\!\!\!\!\!\!\!
G^{+}({\tilde x},{\tilde x'})
=\left(\frac{-\sigma^{2}}{4\pi^2}\right)
\left(\frac{1}{\lambda^2(\tau-\tau'-i \epsilon)^2 
-4\kappa^2\sigma^{-2}\sin^2\left[\sigma(\tau-\tau')/2\right]}\right).
\label{eqn:wgfnrot}
\end{equation}
(It is easy to see that this Wightman function corresponds 
to the case wherein we set $t=(\lambda\tau/\sigma)$, 
$r=(\kappa/\sigma^2)$ and $\Omega=(\sigma^2/\lambda)$ in the 
Feynman propagator~(\ref{eqn:fgfnrot}).) 
The transition probability rate of the rotating detector 
turns out to be non-zero, but the resulting integral cannot 
be expressed in a closed form.
However, it has been evaluated numerically~\cite{letaw81,go83}.
On the other hand, the Bogolubov coefficient~$\beta$ relating 
the modes in the rotating frame and the Minkowski modes vanishes 
identically~\cite{paddy82,letpfa81}.
Also, the kernel corresponding to the operator ${\hat H}$ in the 
rotating frame reduces to~(\ref{eqn:krnl0}) in the coincidence
limit (cf.~(\ref{eqn:rotkrnl})) which then implies that the 
effective Lagrangian reduces to zero in the rotating coordinates 
on regularization.

\subsubsection{In the `cusped' coordinates}

The Wightman function in the Minkowski vacuum evaluated
along the trajectory~(\ref{eqn:cusptraj}) is given by
\begin{equation}
G^{\rm +}({\tilde x},{\tilde x'})
=\left(\frac{-3}{\pi^2}\right)
\left(\frac{1}{12(\tau-\tau'-i\epsilon)^2
+\kappa^2(\tau-\tau')^4}\right).
\end{equation}
(This Wightman function corresponds to the case wherein 
we choose ${\bar x}=(1/2a)$, $a=\kappa$ and ${\bar t}=\tau$
in the Feynman propagator~(\ref{eqn:fgfncusp}).)
On substituting this Wightman function in the 
integral~(\ref{eqn:detrate}), we find that
the resulting transition probability rate of
the detector is given by
\begin{equation}
{\cal R}({\cal E})
=\left(\frac{{\cal E}^2}{8\sqrt{3}\pi^2\kappa^2}\right)\;
\exp-\left(2\sqrt{3}{\cal E}\kappa^{-1}\right).
\end{equation}
However, the Bogolubov coefficient~$\beta$ relating the 
modes in the `cusped' coordinates and the Minkowski modes 
turns out to be zero~\cite{paddy82,letpfa81}.
Also, it is easy to see from Eq.~(\ref{eqn:cuspkrnl}) 
that the kernel in the `cusped' coordinates reduces to the 
kernel~(\ref{eqn:krnl0}) in the coincidence limit.
Therefore, as in the case of the Rindler and the rotating 
coordinates, the effective Lagrangian in the `cusped' 
coordinates vanishes on regularization.

\subsection{Detector response in terms of Bogolubov
coefficients}\label{subsec:bogdet}

It is clear from our discussion in the last section that 
the response of the Unruh-DeWitt detector matches the results
from the Bogolubov transformations only in the case of the 
Rindler coordinates.
In the rotating and the `cusped' coordinate systems, the 
response of the detector turns out to be non-zero even when 
the Bogolubov coefficient~$\beta$ is identically zero.
 
In order to identify the origin of this difference, let us 
now write down the response of a non-inertial Unruh-DeWitt 
detector in terms of the Bogolubov coefficients.
Let $\left\{u_{i}({\tilde x})\right\}$ and 
$\left\{{\bar u}_{k}({\tilde x})\right\}$ 
denote the complete set of positive norm modes 
corresponding to the operator~${\hat H}$ in the 
Minkowski and the non-inertial coordinate systems, 
respectively.
Then, in terms of the modes~$u_{i}({\tilde x})$, the 
Wightman function~(\ref{eqn:wgfn}) in the Minkowski 
vacuum is given by the expression 
\begin{equation}
G^{+}\left[{\tilde x}, {\tilde x'}\right]
=\sum_{i}u_{i}({\tilde x})\,  u_{i}^{*}({\tilde x'}).
\end{equation}
Earlier, we had obtained the Wightman function in the 
non-inertial frame by substituting the trajectory of 
the detector at the two different points ${\tilde x}(\tau)$ 
and ${\tilde x}(\tau')$ in the above expression.
Instead, let us now express the modes~$u_{i}({\tilde x})$
in terms of the modes~${\bar u}_{k}({\tilde x})$ in the 
frame of the detector using the Bogolubov 
transformations~(\ref{eqn:bglbvtrnsfmtn2}).
We obtain that
\begin{eqnarray}
& &\!\!\!\!\!\!\!\!\!\!\!\!\!\!\!\!
\lefteqn{G^{+}\left[{\tilde x}(\tau), 
{\tilde x}(\tau')\right]}\nonumber\\
& &\!\!=\;\sum_{i}\sum_{k}\sum_{l}\;
\left[\alpha_{ki}^{*}\, {\bar u}_{k}({\tilde x})
-\beta_{ki}\, {\bar u}_{k}^{*}({\tilde x})\right]
\left[\alpha_{li}\, {\bar u}_{l}^{*}({\tilde x'})
-\beta_{li}^{*}\, {\bar u}_{l}({\tilde x'})\right]\nonumber\\
& &\!\!=\;\sum_{i}\sum_{k}\sum_{l}\;
\biggl\{\alpha_{ki}^{*}\, \alpha_{li}\,  
{\bar u}_{k}({\tilde x})\, {\bar u}_{l}^{*}({\tilde x'})
-\beta_{ki}\, \alpha_{li}\, {\bar u}_{k}^{*}({\tilde x})\, 
{\bar u}_{l}^{*}({\tilde x'})\nonumber\\
& &\qquad\qquad\qquad\quad
-\;\alpha_{ki}^{*}\, \beta_{li}^{*}\, 
{\bar u}_{k}({\tilde x})\, {\bar u}_{l}({\tilde x'})
+\beta_{ki}\, \beta_{li}^{*}\, {\bar u}_{k}^{*}({\tilde x})\, 
{\bar u}_{l}({\tilde x'})\biggl\}.
\label{eqn:wgfnexp}
\end{eqnarray}
Since we had chosen the trajectory of the detector to be an 
integral curve of a timelike Killing vector field, the modes 
${\bar u}_{k}({\tilde x})$ can be decomposed as follows:
\begin{equation}
{\bar u}_{k}({\tilde x})
= e^{-i\nu_{k}\tau}\; f_{k}({\bar {\bf x}}),
\label{eqn:nimode}
\end{equation} 
where $\tau$ and ${\bar {\bf x}}$ denote the proper time and
the spatial coordinates in the frame of the detector.
Let us now assume that the detector is at the 
position~${\bar {\bf a}}$ in its own coordinate 
system. 
On substituting the modes~(\ref{eqn:nimode}) in the 
expression~(\ref{eqn:wgfnexp}), then substituting 
the resulting Wightman function in Eq.~(\ref{eqn:detprob}) 
and finally integrating over $\tau$ and $\tau'$, we 
find that the transition probability of the detector 
is given by~\cite{go83}
\begin{eqnarray}
& &\!\!\!\!\!\!\!\!\!\!\!\!\!\!\!\!
\lefteqn{{\cal P}({\cal E})
= (2\pi)^2\; \sum_{i}\sum_{k}\sum_{l}\; 
\biggl\{\alpha_{ki}^{*}\, \alpha_{li}\, 
f_{k}({\bar {\bf a}})\,  f_{l}^{*}({\bar {\bf a}})\;
\delta^{(1)}({\cal E}+\nu_{k})\;
\delta^{(1)}({\cal E}+\nu_{l})}\nonumber\\ 
& &\qquad\qquad\qquad\;\;
-\;\beta_{ki}\, \alpha_{li}\,
f_{k}^{*}({\bar {\bf a}})\, f_{l}^{*}({\bar {\bf a}})
\;\delta^{(1)}({\cal E}-\nu_{k})\; 
\delta^{(1)}({\cal E}+\nu_{l})\nonumber\\ 
& &\qquad\qquad\qquad\;\;
-\; \alpha_{ki}^{*}\, \beta_{li}^{*}\, 
f_{k}({\bar {\bf a}})\, f_{l}({\bar {\bf a}})\; 
\delta^{(1)}({\cal E}+\nu_{k})\; 
\delta^{(1)}({\cal E}-\nu_{l})\nonumber\\ 
& &\qquad\qquad\qquad\;\;
+\; \beta_{ki}\, \beta_{li}^{*}\,
f_{k}^{*}({\bar {\bf a}})\,  f_{l}({\bar {\bf a}})\;
\delta^{(1)}({\cal E}-\nu_{k})\;
 \delta^{(1)}({\cal E}-\nu_{l})\biggl\}.
\label{eqn:detbog}
\end{eqnarray}

Recall the fact that the modes ${\bar u}_{k}({\tilde x})$ 
are positive {\it norm}\/ modes. 
Let us now assume that the definition of positive norm 
modes match the definition of positive frequency modes 
in the frame of the detector for all frequencies (i.e. 
$\nu_{k}\ge 0$ $\forall k$).
In such a situation, only the last term in the expression 
above will contribute to ${\cal P}({\cal E})$ with the 
result
\begin{equation}
{\cal P}({\cal E})
=(2\pi)^2 \, \vert f_{\cal E}({\bar {\bf a}})\vert^2\;\,
\sum_{i} \vert \beta_{{\cal E} i}\vert^2.\label{eqn:comp}
\end{equation} 
Clearly, in such cases, the detector response will prove 
to be non-zero only when the Bogolubov coefficient~$\beta$ 
is not zero.
Moreover, the detector response will actually match 
the expectation value of the number operator in the 
non-inertial frame evaluated in the Minkowski vacuum 
(compare Eq.~(\ref{eqn:comp}) 
above with Eq.~(\ref{eqn:expval})). 
This is exactly what happens in the case of the Rindler 
coordinates.

On the other hand, if some of the negative frequency modes 
in the frame of the detector have a positive norm (i.e. 
$\nu_{k}<0$ for some values of $k$), then it is easy 
to see from Eq.~(\ref{eqn:detbog}) that the first term can 
contribute to ${\cal P}({\cal E})$ even when the Bogolubov 
coefficient $\beta$ turns out to be zero.
In such a case, the transition probability of the 
non-inertial detector reduces to 
\begin{equation}
{\cal P}({\cal E})
=(2\pi)^2\, \vert f_{-{\cal E}}({\bar {\bf a}})\vert^2\;\,
\sum_{i} \vert \alpha_{-{\cal E} i}\vert^2.
\end{equation}
It is known that there exists a range of frequencies for 
which negative frequency modes have a positive norm in
the rotating as well as the `cusped' 
coordinates~\cite{letpfa81}.
It is these modes that excite the detector as a result 
of which the response of the Unruh-DeWitt detector along 
these trajectories proves to be non-zero even when the 
Bogolubov coefficient $\beta$ is identically zero.

We had pointed out earlier that it is the norm of the 
modes (rather than their frequency) that has to be taken 
into account in decomposing the quantum field in terms 
of the creation and the annihilation operators. 
These operators in turn define the vacuum state of the 
field.
Our discussion above points to the fact that non-trivial 
effects can arise in the vacuum (even in situations wherein
the Bogolubov coefficient~$\beta$ proves to be zero) when 
the norm and the frequency of the modes do not match.
Though such modes arise in flat spacetime due to the 
non-inertial motion of the observer, these modes occur 
even in {\it inertial}\/ frames in backgrounds such as 
a time-independent electric field. 
As we shall see later, it is these modes that turn out to 
be responsible for exciting an inertial detector in such
a background (under conditions wherein particle production 
is expected to occur). 

\section{In the presence of boundaries in flat 
spacetime}\label{sec:bndrs}

In this section, we shall consider the response of inertial 
and rotating Unruh-DeWitt detectors when boundaries are present 
in flat spacetime.
We shall discuss two cases: (i)~the response of an inertial 
detector in the Casimir vacuum and (ii)~the response of a 
rotating detector when boundary conditions are imposed on 
the field at the horizon in the rotating frame. 
We shall compare the response of these detectors with the 
results from the effective Lagrangian approach.
The system we shall consider here is a massless scalar 
field~$\Phi$ described by the action~(\ref{eqn:gravact}).

\subsection{In an inertial frame}\label{subsec:casine}

Let us first consider the response of an inertial detector
in the Casimir vacuum.
Let us impose periodic boundary conditions on the quantum 
field~${\hat \Phi}$ along the $x$-axis. 
In other words, we shall assume that the field takes on the 
same value at, say, $x$ and $(x+L)$.
In such a case, the positive norm modes of the quantum field 
are given by
\begin{equation}
u_{\bf k}(t,{\bf x})
=\left(\frac{1}{\sqrt{(2\pi)^2\, 2\omega_k L}}\right)\;
e^{-i\omega_k t}\; e^{i{\bf k}\cdot{\bf x}},
\end{equation}
where $\omega_k=\vert{\bf k}\vert$, $k_x=(2n\pi/L)$ and 
$n=0,\pm 1,\pm 2,\ldots$.
Now, consider an inertial detector stationed at a point, 
say, ${\bf a}$. 
The transition amplitude~${\cal A}_{\bf k}^{*}({\cal E})$ of 
such a detector in the Casimir vacuum is proportional to a 
delta function as in Eq.~(\ref{eqn:detampine}).
Since~$\omega_k\ge 0$ for all ${\bf k}$, an inertial detector 
does not respond in the Casimir vacuum for the same reasons an 
inertial detector does not respond in the Minkowski vacuum.

On the other hand, it is easy to show that the 
effective Lagrangian proves to be non-zero in 
such a situation~\cite{sriram98}. 
The operator~${\hat H}$ in such a case corresponds to that 
of a free particle along the $t$, $y$ and $z$ directions.
Whereas, along the $x$-direction, the eigen functions of 
the operator~${\hat H}$ should be assumed to take on the 
same value at $x$ and $(x+L)$.
Therefore, the kernel in such a case can be written as
\begin{equation}
K({\tilde x}, {\tilde x}; s) 
=\left(\frac{i}{(4\pi i s)^{3/2}}\right)\, 
\langle x\vert e^{-i{\hat H'}s}\vert x\rangle,\quad
{\rm where}\quad {\hat H'}= -d_x^2 .
\end{equation}
On imposing the periodic boundary condition, the normalized 
eigen functions of the operator ${\hat H'}$ corresponding to 
an energy eigen value $E=(4n^2 \pi^2 /L^2)$ are given by
\begin{equation}
\Psi_E(x) 
= \left({\frac{1}{\sqrt{L}}}\right)\, e^{(2i n \pi x/L)},
\quad{\rm where}\quad n=0, \pm 1, \pm 2, \ldots.
\end{equation}
The corresponding kernel in the coincidence limit can then 
be written using the Feynman-Kac formula as follows (see, 
for instance, Ref.~\cite{fh65}, p.~88):
\begin{equation}
\langle x\vert e^{-i{\hat H'} s}\vert x\rangle
=\left(\frac{1}{L}\right) \sum_{n=-\infty}^{\infty}
\exp-\left(4in^2 \pi^2 s/L^2\right).
\end{equation}
Using the Poisson sum formula, this sum can be rewritten 
as (cf.~Ref.~\cite{mf53}, p.~483):
\begin{equation}
\langle x\vert e^{-i{\hat H'}s}\vert x\rangle
=\left(\frac{1}{\sqrt{4\pi i s}}\right) 
\sum_{n=-\infty}^{\infty} \exp \left(in^2 L^2/4s\right).
\end{equation}
Therefore, the complete kernel is given by
\begin{eqnarray} 
K({\tilde x}, {\tilde x}; s) 
&=& \left(\frac{1}{16 \pi^2 i s^2}\right)
\sum_{n=-\infty}^{\infty} 
\exp\left(in^2L^2/4s\right)\nonumber\\
&=& \left(\frac{1}{16 \pi^2 i s^2}\right)
\left\{1+2\sum_{n=1}^{\infty} 
\exp\left(in^2L^2/4s\right)\right\}.
\label{eqn:caskrnl}
\end{eqnarray}
On substituting this kernel in Eq.~(\ref{eqn:lcorr}) 
and subtracting the quantity ${\cal L}_{\rm corr}^0$, 
we obtain that~\cite{sriram98}
\begin{equation}
{\bar {\cal L}}_{\rm corr}
=\left(\frac{1}{\pi^2\, L^4}\right)
\sum_{n=1}^{\infty} n^{-4}
= \left(\frac{1}{\pi^2\, L^4}\right)\;\zeta(4) 
=\left(\frac{\pi^2}{90\,L^4}\right),
\end{equation}
where we have made use of the fact that $\zeta(4)=(\pi^4/90)$   
(cf.~Ref.~\cite{arfken85}, p.~334).
Clearly, this effective Lagrangian is a real quantity 
and, in fact, corresponds to the Casimir energy arising 
due to the boundaries (see, for e.g., Ref.~\cite{iz80}, 
pp.~138--142).

\subsection{In a rotating frame}\label{subsec:casrot}

In Section~\ref{sec:nifrms}, we had found that a  detector 
in a rotating frame responds non-trivially in the Minkowski 
vacuum.
We had also shown that it is the negative frequency modes which 
have a positive norm that are responsible for exciting the 
rotating detector.
It is easy to see from the line element~(\ref{eqn:rotmet}) 
that the velocity of a observer stationed at a radius~$r$ 
greater than $\Omega^{-1}$ in the rotating frame exceeds 
the velocity of light.
In other words, flat spacetime exhibits a horizon in the 
rotating frame at $r=\Omega^{-1}$.
Due to this reason, it has been argued in literature that 
the quantum field has to be assumed to vanish on the horizon.
Interestingly, imposing such a boundary condition at the 
horizon leads to a situation wherein there exists no negative 
frequency modes with a positive norm in the rotating frame 
and, as a result, the rotating detector ceases to 
respond~\cite{tevian96}.

Two important points need to be noted about this curious result.
Firstly, imposing a boundary condition at the horizon alters 
the vacuum structure of the field and, hence, the field is not 
any more in the Minkowski vacuum but is in a Casimir vacuum.
Secondly, we had seen earlier that the effective Lagrangian 
vanishes in the rotating frame.
But, if we impose a boundary condition on the field at a
particular radius, the effective Lagrangian for such a case 
would turn out to be non-zero and would, in fact, correspond 
to the Casimir energy of a cylinder (see, for e.g., 
Ref.~\cite{cylinder}).

\section{In classical electromagnetic 
backgrounds}\label{sec:embckgnds}

The quantum field we shall consider in this section is a 
{\it complex}\/ scalar field~$\Phi$ described by the action
\begin{equation}
S[\Phi]=\int d^4x 
\left\{\left(\partial_{\mu}\Phi+iqA_{\mu}\Phi\right)
\left(\partial^{\mu}\Phi^*-iqA^{\mu}\Phi^*\right) 
- m^2\Phi\Phi^*\right\},\label{eqn:emact}
\end{equation}
where $A^{\mu}$ is the vector potential describing the 
classical electromagnetic background and $q$ and $m$ 
are the charge and the mass of a single quanta of the 
scalar field.
Varying this action leads to an equation of motion such as 
Eq.~(\ref{eqn:mtn}) with the operator ${\hat H}$ given by
\begin{equation}
{\hat H}\equiv \left(\partial_{\mu}+iqA_{\mu}\right)
\left(\partial^{\mu}+iqA^{\mu}\right).\label{eqn:emop}
\end{equation}

\subsection{The non-linearly coupled
detector}\label{subsec:nlcd}

The Lagrangian~(\ref{eqn:lint}) describes the interaction 
between the Unruh-DeWitt detector and a {\it real}\/ scalar 
field.
For the case of the complex scalar field we are considering 
here, the interaction Lagrangian~(\ref{eqn:lint}) can be 
generalized to
\begin{equation}
{\cal L}_{\rm int}
={\bar c}\,
\,\biggl(\mu(\tau)\, \Phi[{\tilde x}(\tau)]
+ \mu^*(\tau)\, \Phi^*[{\tilde x}(\tau)]\biggl).
\label{eqn:lintc}
\end{equation}
Under a gauge transformation of the form: $A^{\mu} \to 
\left(A^{\mu}+\partial^{\mu}\chi\right)$, the complex scalar 
field transforms as: $\Phi\to \left(\Phi\, e^{-iq\chi}\right)$. 
Clearly, the interaction Lagrangian~(\ref{eqn:lintc}) will 
not be invariant under such a gauge transformation, unless 
we assume that the monopole moment transforms as follows:
$\mu\to \left(\mu\, e^{iq\chi}\right)$. 
However, we would like to treat the detector part of 
the coupling, viz.~the monopole moment $\mu(\tau)$, 
as a quantity that transforms as a scalar under gauge 
transformations. 
In such a case, the simplest of the Lagrangians that 
is explicitly gauge-invariant is the non-linear 
interaction~\cite{sriram99}
\begin{equation}
{\cal L}_{\rm int} = {\bar c}\, \mu(\tau)\,
\biggl(\Phi[{\tilde x}(\tau)]\,
\Phi^*[{\tilde x}(\tau)]\biggl).\label{eqn:giint}
\end{equation}
It is important to note that demanding gauge invariance 
naturally leads to non-linear interactions.
A physical manifestation of gauge invariance is charge
conservation.
As we shall see later, the non-linear and gauge-invariant
interaction~(\ref{eqn:giint}) leads to the excitation of a
particle-anti-particle pair thereby conserving charge.

In an electromagnetic background, the quantized complex 
scalar field ${\hat \Phi}$ can, in general, be decomposed 
as follows (see, for instance, Ref.~\cite{manogue88}):
\begin{equation}
{\hat \Phi}({\tilde x})=\sum_i 
\left[{\hat a}_i\, u_i({\tilde x}) 
+{\hat b}_{i}^{\dag}\, v_{i}({\tilde x})\right],
\label{eqn:decompc}
\end{equation}
where $u_{i}({\tilde x})$ and $v_{i}({\tilde x})$ are positive 
and negative {\it norm}\/ modes, respectively\footnote{The only 
non-trivial commutation relations satisfied by the two sets of 
operators $\left\{{\hat a}_{i},{\hat a}_{i}^{\dag}\right\}$ 
and $\left\{{\hat b}_{i},{\hat b}_{i}^{\dag}\right\}$ are: 
$\left[{\hat a}_{i},{\hat a}_{j}^{\dag}\right]
=\left[{\hat b}_{i}, {\hat b}_{j}^{\dag}\right]
=\delta_{ij}$. All other commutators vanish.}.
These modes are normalized with respect to the following 
gauge-invariant scalar product (see, for e.g., 
Ref.~\cite{fulling89}, p.~227)
\begin{equation}
(u_{i},u_{j})
=-i\int\limits_{t=0}^{} d^3x\, \left(u_{i}
\left[\partial_t-iqA_t\right]u_{j}^*
-u_{j}^* \left[\partial_t+iqA_t\right]u_{i}\right),
\label{eqn:sclrprdct}
\end{equation}
where~$A_t$ is the zeroth component of the vector potential
$A^{\mu}$.
The vacuum state $\vert 0\rangle$ of the quantum 
field~${\hat \Phi}$ is defined as the state that 
is annihilated by both the operators ${\hat a}_i$
and ${\hat b}_i$ for all~$i$.

Let us now assume that the quantized complex 
scalar field~${\hat \Phi}$ is initially in the
vacuum state~$\vert 0\rangle$. 
Then, up to the first order in perturbation theory, 
the amplitude of transition of the detector that is 
coupled to the field through the interaction 
Lagrangian~(\ref{eqn:giint}) is given by
\begin{eqnarray}
\!\!\!\!\!\!\!\!\!\!
{\tilde {\cal A}}({\cal E}) 
&=&\left(\frac{1}{2}\right)\,
\int\limits_{-\infty}^{\infty} d\tau\, e^{i {\cal E}\tau}\, 
\langle\Psi\vert\biggl({\hat \Phi}[{\tilde x}(\tau)]\,
{\hat \Phi}^{\dag}[{\tilde x}(\tau)]\nonumber\\
& &\qquad\qquad\qquad\qquad\qquad\qquad\qquad
+\;{\hat \Phi}^{\dag}[{\tilde x}(\tau)]\,
{\hat \Phi}[{\tilde x}(\tau)]\biggl)\vert 0\rangle,
\label{eqn:detampnld}
\end{eqnarray}
where, as in the case of the Unruh-DeWitt detector, 
${\cal E}=({\bar E}-{\bar E}_0)$, ${\bar E}_0$ and 
${\bar E}$ are the energy eigen values corresponding 
to the ground state and the excited state of the 
detector and $\vert \Psi \rangle$ is the state of the 
quantum field after its interaction with the detector.
On substituting the decomposition~(\ref{eqn:decompc}) 
for the field ${\hat \Phi}$ in the transition 
amplitude~(\ref{eqn:detampnld}), we obtain that
\begin{eqnarray}
\!\!\!\!\!\!\!\!\!\!
{\tilde {\cal A}}^*({\cal E}) 
&=& \int\limits_{-\infty}^{\infty} d\tau\, e^{-i {\cal E}\tau}\, 
\biggl\{{\tilde G}_{1}[{\tilde x}(\tau),{\tilde x}(\tau)]\;
\langle 0\vert\Psi\rangle\nonumber\\
& &\qquad\qquad\qquad\quad
+\; \sum_{i}\sum_{j}
u_{i}[{\tilde x}(\tau)]\,  v_{j}^*[{\tilde x}(\tau)]\; 
\langle 0\vert {\hat a}_{i}{\hat b}_{j}
\vert\Psi\rangle\biggl\},\label{eqn:transamp1}
\end{eqnarray}
where ${\tilde G}_{1}[{\tilde x},{\tilde x'}]$ is the two-point 
function defined as
\begin{equation}
{\tilde G}_{1}\left[{\tilde x}, {\tilde x'}\right]
=\left(\frac{1}{2}\right)
\langle 0 \vert 
\left[{\hat \Phi}({\tilde x})\,
{\hat \Phi}^{\dag}({\tilde x'})\,
+{\hat \Phi}^{\dag}({\tilde x'})\,
{\hat \Phi}({\tilde x})\right]
\vert 0 \rangle.
\end{equation}
This two-point function can be expressed in the terms of the 
modes $u_{i}({\tilde x})$ and $v_{i}({\tilde x})$ as follows:
\begin{equation}
{\tilde G}_{1}\left[{\tilde x}, {\tilde x'}\right]
=\left(\frac{1}{2}\right)\,
\sum_{i}\left[u_{i}({\tilde x})\, u_{i}^*({\tilde x'})
+v_{i}({\tilde x})\, v_{i}^*({\tilde x'})\right].
\end{equation}
The first term in the transition amplitude~(\ref{eqn:transamp1}) 
contributes even when $\vert \Psi\rangle=\vert 0 \rangle$.
But, since the two-point function ${\tilde G}_{1}[{\tilde x},
{\tilde x}]$ is an infinite quantity, we shall hereafter drop 
this term and assume that the transition amplitude 
${\tilde {\cal A}}^*({\cal E})$ above is given only by the 
second term\footnote{We can formally justify this procedure 
by saying that we are normal ordering the creation and the 
annihilation operators in the matrix element in the transition 
amplitude~(\ref{eqn:detampnld}). 
The divergent first term in Eq.~(\ref{eqn:transamp1}) would not 
arise in such a case.}. 
The second term contributes only when $\vert \Psi\rangle= 
{\hat a}_{i}^{\dag}{\hat b}_{j}^{\dag}\vert 0\rangle
=\vert 1_{i}, 1_{j}\rangle$. 
This implies that the interaction of the field with the detector 
leads to the excitation of a particle-anti-particle pair.
Since the quantum field we are considering here is a 
{\it charged}\/ scalar field, the excitation of such 
a pair in essential for charge conservation.
As we had pointed out above, it is the non-linear and 
gauge-invariant nature of the interaction 
Lagrangian~(\ref{eqn:giint}) that ensures that such a 
pair is indeed excited. 

Let us now consider the response of an inertial detector stationed
at a point ${\bf a}$ in the Minkowski vacuum.
In the absence of an electromagnetic background, the positive 
and negative norm modes are related as follows: $v_{i}({\tilde x})
=u_{i}^*({\tilde x})$.
Moreover, as we have pointed out earlier, the definition of positive 
norm modes match the definition of positive frequency modes in the 
Minkowski coordinates. 
It is then clear from Eq.~(\ref{eqn:transamp1}) that it is only 
the positive frequency modes $u_{i}({\tilde x})$ that contribute 
to the transition amplitude ${\tilde {\cal A}}^*({\cal E})$ in 
such a situation.
Therefore, the transition amplitude of the detector 
corresponding to a pair of modes, say, ${\bf k}$ and 
${\bf l}$ of the quantum field  is given by
\begin{equation}
{\tilde {\cal A}}_{{\bf k},{\bf l}}^*({\cal E})
=\left(\frac{e^{i({\bf k}+{\bf l})\cdot{\bf a}}}{\sqrt{(2\pi)^4\, 
4\omega_k\omega_l}}\right)\; \delta^{(1)}({\cal E}+
\omega_k+\omega_l),\label{eqn:minkampnld}
\end{equation}
where, for a given mode ${\bf k}$,
$\omega_{k}=\left(\vert {\bf k}\vert^2+m^2\right)^{1/2}$.
The quantities $\omega_k$ and $\omega_l$ are always $\ge m$
and, since ${\cal E}>0$ as well, the argument of the delta 
function above is a positive definite quantity and, hence, 
the transition 
amplitude~${\tilde {\cal A}}_{{\bf k},{\bf l}}^*({\cal E})$ 
reduces to zero for {\it all}\/ ${\bf k}$ and ${\bf l}$. 
In other words, just like the Unruh-DeWitt detector, the 
non-linearly coupled detector does not respond in the 
Minkowski vacuum when in inertial motion.

The transition probability of the detector to all possible 
final states $\vert \Psi\rangle$ of the field is given by 
the expression
\begin{equation}
{\tilde {\cal P}}({\cal E}) 
=\sum_{\vert\Psi\rangle}{\vert {\tilde {\cal A}}({\cal E})\vert}^2
= \int\limits_{-\infty}^\infty d\tau\, 
\int\limits_{-\infty}^\infty d\tau'\, 
e^{-i{\cal E}(\tau-\tau')}\, 
{\tilde {\cal G}}\left[{\tilde x}(\tau), {\tilde x}(\tau')\right],
\label{eqn:detprobnld}
\end{equation}
where ${\tilde {\cal G}}\left[{\tilde x}(\tau), 
{\tilde x}(\tau')\right]$ is a four-point function 
defined as
\begin{eqnarray}
& &\!\!\!\!\!\!\!\!\!\!\!\!
\lefteqn{{\tilde {\cal G}}\left[{\tilde x}(\tau), 
{\tilde x}(\tau')\right]}\nonumber\\
& &=\;\left(\frac{1}{4}\right)\,\langle 0 \vert 
\left({\hat \Phi}[{\tilde x}(\tau)]\,
{\hat \Phi}^{\dag}[{\tilde x}(\tau)]
+{\hat \Phi}^{\dag}[{\tilde x}(\tau)]\,
{\hat \Phi}[{\tilde x}(\tau)]\right)\nonumber\\
& &\qquad\qquad\qquad\;\times\;
\left({\hat \Phi}[{\tilde x}(\tau')]\,
{\hat \Phi}^{\dag}[{\tilde x}(\tau')]
+{\hat \Phi}^{\dag}[{\tilde x}(\tau')]\,
{\hat \Phi}[{\tilde x}(\tau')]\right)
\vert 0 \rangle.\label{eqn:fptfn}
\end{eqnarray}
Using the decomposition~(\ref{eqn:decompc}), this four-point 
function can be expressed as follows:
\begin{equation}
\!\!\!\!\!\!\!\!\!\!\!\!
{\tilde {\cal G}}[{\tilde x},{\tilde x'}]
={\tilde G}_{1}[{\tilde x},{\tilde x}]\; 
{\tilde G}_{1}[{\tilde x'},{\tilde x'}]
+\sum_i \left[u_i({\tilde x})\, u_i^*({\tilde x'})\right]\;
\sum_j \left[v_j^*({\tilde x})\, v_j({\tilde x'})\right].
\label{eqn:fptfn1}
\end{equation}
The first term in this expression is a product of two two-point 
functions evaluated at the {\it same}\/ spacetime point and hence 
is an infinite quantity\footnote{It should be pointed out here 
that this term would not arise had we normal ordered the creation 
and the annihilation operators in the matrix element in the 
transition amplitude~(\ref{eqn:detampnld}).}.
Therefore, we shall drop this term and assume that the 
four-point function ${\tilde {\cal G}}[{\tilde x},{\tilde x'}]$ 
above is given only by the second term.

We had pointed out above that, in the absence of an 
electromagnetic background, the positive and the negative
norm modes are related by the following expression:
$v_{i}({\tilde x}) =u_{i}^*({\tilde x})$.
It is then useful to note that, in such a case, the 
second term in the four-point function 
${\tilde {\cal G}}[{\tilde x},{\tilde x'}]$ above will be given 
by the square of the Wightman function in the Minkowski vacuum. 
Therefore, when in inertial motion, the transition probability 
rate of the non-linearly coupled detector in the Minkowski vacuum 
would be identically zero (for exactly the same reasons) as it 
is in the case of the Unruh-DeWitt detector.

In the following three sections, we shall study the response
of the non-linearly coupled detector in:~(i)~a time-dependent 
electric field, (ii)~a time-independent electric field and 
(iii)~a time-independent magnetic field, backgrounds. 
We had seen earlier that detectors on non-inertial trajectories 
respond non-trivially even in the Minkowski vacuum. 
Therefore, in order to avoid the effects due to non-inertial motion 
and to isolate the effects that arise due to the electromagnetic 
background, we shall restrict our attention to {\it inertial}\/ 
trajectories here. 
We shall compare the response of the inertial detector with 
the results expected from the Bogolubov transformations and 
the effective Lagrangian approach.

\subsection{In time-dependent electric field 
backgrounds}\label{subsec:tdef}

A time-dependent electric field background can be described 
by following vector potential:
\begin{equation}
A^{\mu}=(0, A(t), 0,0),\label{eqn:tde}
\end{equation}
where $A(t)$ is an arbitrary function of $t$.
This vector potential gives rise to the electric field 
${\bf E}=-(dA/dt)\, {\hat {\bf x}}$, where ${\hat {\bf x}}$ 
is the unit vector along the positive $x$-direction.
The modes of a quantum field evolving in such a time-dependent 
electric field background can be decomposed as
\begin{equation}
u_{\bf k}(t,{\bf x})
=g_{\bf k}(t)\; e^{i{\bf k}\cdot{\bf x}}.
\label{eqn:modetde}
\end{equation}
In general, modes at early and late times will be related by 
a non-zero Bogolubov coefficient $\beta$ and the expectation 
value of the number operator (corresponding to a given mode
of the quantum field) at late times in the in-vacuum 
will be given by Eq.~(\ref{eqn:expval}) (see, for e.g., 
Ref.~\cite{fgs91}, Sec.~2.1).

Now, consider a detector that is stationed at a particular
point.
Along the world line of such a detector, the second term
in four-point function~(\ref{eqn:fptfn1}) corresponding 
to the modes~(\ref{eqn:modetde}) is given by
\begin{equation}
{\tilde {\cal G}}(t,t')
=\sum_{\bf k} \sum_{\bf l} \left[g_{\bf k}(t)\, 
g_{\bf l}(t)\, g_{\bf k}^*(t')\, g_{\bf l}^*(t')\right]
\end{equation}
and the transition probability of the detector reduces to
\begin{equation}
{\tilde {\cal P}}({\cal E})
=\sum_{\bf k}\sum_{\bf l} \vert g_{\bf kl}({\cal E})\vert^2,
\end{equation}
where
\begin{equation}
g_{\bf kl}({\cal E})
=\int\limits_{-\infty}^{\infty} dt\;
e^{-i{\cal E}t}\, \left[g_{\bf k}(t)\, g_{\bf l}(t)\right].
\end{equation}
Clearly, the response of the inertial detector will, in
general, be non-zero.

Let us now assume that: (i) the function $A(t)$ behaves such 
that the electric field vanishes in the past and future 
infinity, (ii) the detector is switched on 
for a finite time interval in the future asymptotic 
domain and  (iii) the effects that arise due 
to switching~\cite{ftdet1,ftdet2,ftdet3} can be neglected.
Then, by relating the modes at future and past infinity 
using the Bogolubov transformations, we can express the 
response of the detector (in the in-vacuum) in terms 
of the Bogolubov coefficients (as we have done in 
Section~\ref{subsec:bogdet}).
We find that the transition probability of the detector
is given by~\cite{sriram99}:
\begin{eqnarray}
{\tilde {\cal P}}({\cal E})
&=&(2\pi)^2\;\sum_{\bf k}\sum_{\bf l}
\biggl(\vert \alpha_{\bf k}\vert^2\,
\vert \beta_{\bf l}\vert^2\;
\delta^{(2)}\left({\cal E}+\omega_{-}\right)\nonumber\\
& &\qquad\qquad\qquad\;+2\, \left[{\rm Re.}
(\alpha_{\bf k}\alpha_{\bf l}^*
\beta_{\bf k}^* \beta_{\bf l})\right]\;
\delta^{(1)}\left({\cal E}+\omega_{-}\right)\;
\delta^{(1)}\left({\cal E}-\omega_{-}\right)\nonumber\\
& &\qquad\qquad\qquad\;+\, 4\, \vert \beta_{\bf k}\vert^2\,
\left[{\rm Re.}(\alpha_{\bf l}\beta_{\bf l}^*)\right]\; 
\delta^{(1)}\left({\cal E}-\omega_{+}\right)\; 
\delta^{(1)}\left({\cal E}-\omega_{-}\right)\nonumber\\
& &\qquad\qquad\qquad\;
+\, \vert \beta_{\bf k}\vert^2\;
\vert \alpha_{\bf l}\vert^2\;
\delta^{(2)}\left({\cal E}-\omega_{-}\right)\nonumber\\
& &\qquad\qquad\qquad\;
+\,\vert \beta_{\bf k}\vert^2\; 
\vert \beta_{\bf l}\vert^2\;
\delta^{(2)}\left({\cal E}-\omega_{+}\right)\biggl),
\end{eqnarray}
where $\omega_{\pm}=(\omega_ {\bf k}\pm\omega_{\bf l})$, with 
$\omega_{\bf k}$ and $\omega_{\bf l}$ being the positive 
definite (in fact $\ge m$) frequencies corresponding to the 
modes ${\bf k}$ and ${\bf l}$ in the out-region.
Clearly, the detector responds {\it only}\/ when the
Bogolubov coefficient~$\beta$ is non-zero (i.e. only 
when particle production takes place).
However, it is evident that the transition probability rate 
of detector we have obtained above is {\it not}\/ proportional 
to the number of particles produced by the time-dependent
electric field background.

The feature that the response of the detector does not turn
out to  be proportional to the number of particles produced 
by the background should not come as a surprise and, in fact, 
it can be attributed to the non-linearity of the interaction 
Lagrangian~(\ref{eqn:giint}) for the following two reasons. 
Firstly, it can be easily shown that in a time-dependent 
gravitational background with asymptotically static domains, 
the response of the Unruh-DeWitt detector in the out-region 
will be given by an expression such as Eq.~(\ref{eqn:comp}).
In other words, the response of the Unruh-DeWitt detector in 
such a situation will be proportional to the number of particles 
produced by the background (cf.~Ref.~\cite{bd82}, pp.~57-59).
Secondly, it is known that the response of a detector that 
is coupled to the stress-energy tensor of the quantum field 
(which is evidently a non-linear interaction) does {\it not}\/ 
reflect the particle content of the field~\cite{ps87}.
As we have discussed earlier, demanding gauge invariance 
naturally leads to non-linear interaction Lagrangians. 
Therefore, quite generically, we can expect that the 
response of detectors in classical electromagnetic 
backgrounds will {\it not}\/ be proportional to the 
amount of particles produced by the background.

The imaginary part of the effective Lagrangian for a 
time-dependent electric field background is, in general,
expected to be non-zero implying that such backgrounds
always produce particles.
However, it should be added that evaluating the effective 
Lagrangian for an arbitrary time-dependent electric field 
proves to be a difficult task and the effective Lagrangian
has been obtained in a closed form only in a few cases (for 
efforts on evaluating the effective Lagrangian for non-trivial 
backgrounds, see Ref.~\cite{dh98} and references therein).

\newpage
\subsection{In time-independent electric field backgrounds}
\label{subsec:tindef}

Consider the vector potential
\begin{equation}
A^{\mu}=(A(x), 0, 0,0),\label{eqn:tine}
\end{equation}
where $A(x)$ is an arbitrary function of $x$.
Such a vector potential gives rise to a time-independent 
electric field along the $x$-direction given by ${\bf E}
=-(dA/dx)\,{\hat {\bf x}}$.
In such a case, the modes of the quantum field~${\hat \Phi}$ 
can be decomposed as follows:
\begin{equation}
u_{\omega\bf k_{\perp}}(t,{\bf x})
=e^{-i\omega t}\, f_{\omega{\bf k}_{\perp}}(x)\, 
e^{i{\bf k}_{\perp}\cdot{\bf x}_{\perp}},
\label{eqn:modetine}
\end{equation}
where ${\bf k}_{\perp}$ is the wave vector along the 
perpendicular direction.
Due to lack of time dependence, the Bogolubov coefficient 
$\beta$ relating these modes at two different times is 
trivially zero.
Though the Bogolubov coefficient $\beta$ is zero, particle 
production takes place in such backgrounds due to a totally 
different phenomenon.
It is well known that if the depth of the potential 
$\left[q A(x)\right]$ is greater than $(2m)$, then 
the corresponding electric field will produce particles 
due to Klein paradox (see Ref.~\cite{manogue88} and 
references therein; also see Ref.~\cite{cd99} for a
recent discussion). 
It is then interesting to examine whether an inertial detector 
in a time-independent electric field background will respond 
under the same condition.

Consider a detector that is stationed at a particular point.
It is easy to see from the form of the 
modes~(\ref{eqn:modetine}) that the transition 
amplitude~${\tilde {\cal A}}_{{\bf k},{\bf l}}^{*}({\cal E})$ 
of such a detector will be proportional to a delta function 
as in the case of an inertial detector in the 
Minkowski vacuum (cf.~Eq.~(\ref{eqn:minkampnld})).
But, unlike the Minkowski case wherein the definition of
positive frequency modes match the definition of positive
norm modes, in a time-independent electric field background, 
there exist negative frequency modes which have a positive 
norm whenever the depth of the potential $\left[q A(x)\right]$
is greater than $(2m)$.
In other words, when Klein paradox occurs in an electric field 
background, $\omega_{k}$ and $\omega_{l}$ appearing in the 
argument of the delta function in Eq.~(\ref{eqn:minkampnld}) can 
be negative and, hence, there exists a range of values of these 
two quantities for which this argument can be zero.
These modes excite the detector as a result of which the 
response of an inertial detector proves to be non-zero in 
such a background. 

We shall now show (for the special case of the step potential) 
as to how there exist negative frequency modes which have a 
positive norm when the depth of the potential 
$\left[q A(x)\right]$ is greater than $(2m)$.
In order to show that, let us evaluate the norm of the mode
$u_{\omega{\bf k}_{\perp}}(t,{\bf x})$. 
On substituting the mode~(\ref{eqn:modetine}) and 
the vector potential~(\ref{eqn:tine}) in the scalar 
product~(\ref{eqn:sclrprdct}), we obtain that
\begin{equation}
(u_{\omega{\bf k}_{\perp}},u_{\omega{\bf k}_{\perp}})
=2\, (2\pi)^2\, \delta^{(2)}(0) 
\int\limits_{-\infty}^{\infty} dx\,
\left[\omega-qA(x)\right]\,
\vert f_{\omega{\bf k}_{\perp}}(x)\vert^2.
\label{eqn:sp}
\end{equation} 
Let us now assume that $A(x)=-\left(\Theta(x)\,V\right)$, 
where $\Theta(x)$ is the step-function and $V$ is a constant. 
For such a case, the function $f_{\omega{\bf k}_{\perp}}$ 
is given by
\begin{equation}
f_{\omega{\bf k}_{\perp}}(x)
=\Theta(-x)\left(e^{ik_L x} + R_{\omega{\bf k}_{\perp}}\, 
e^{-ik_L x}\right)
+\Theta(x)\,T_{\omega{\bf k}_{\perp}}\, e^{ik_R x},
\end{equation}
where 
\begin{equation}
k_R=\left[(\omega+qV)^2-\vert{\bf k}_{\perp}\vert^2
-m^2\right]^{1/2}\quad{\rm and}\quad
k_L=\left[\omega^2-\vert{\bf k}_{\perp}\vert^2
-m^2\right]^{1/2}.
\end{equation}
The quantities $R_{\omega{\bf k}_{\perp}}$ and 
$T_{\omega{\bf k}_{\perp}}$ are the usual reflection 
and tunneling amplitudes.
They are given by the expressions
\begin{equation}
R_{\omega{\bf k}_{\perp}}
=\left(\frac{k_L-k_R}{k_L+k_R}\right)\quad
{\rm and}
\quad T_{\omega{\bf k}_{\perp}}
=\left(\frac{2k_L}{k_L+k_R}\right).
\end{equation}
If we now assume that $k_R$ and $k_L$ are real quantities, 
then it is easy to show that, for the case of the step 
potential we are considering here, the scalar 
product~(\ref{eqn:sp}) is given by
\begin{equation}
(u_{\omega{\bf k}_{\perp}},u_{\omega{\bf k}_{\perp}})
=(2\pi)^3\, \delta^{(3)}(0)\, 
\left[\omega \left(1+R_{\omega{\bf k}_{\perp}}^2\right)
+(\omega+qV)\, T_{\omega{\bf k}_{\perp}}^2\right].
\label{eqn:sp1}
\end{equation} 
Let us now set ${\bf k}_{\perp}=0$.
Also, let us assume that $\omega =-(m+\varepsilon)$ and 
$(qV)=(2m+\varepsilon)$, where $\varepsilon$ is a positive 
definite quantity.
For such a case, $R_{\omega0}=1$, $T_{\omega0}=2$ and
the scalar product~(\ref{eqn:sp1}) reduces to
\begin{equation}
(u_{\omega0},u_{\omega0})
=2\, (m-\varepsilon)\; (2\pi)^3\;  \delta^{(3)}(0)
\end{equation}
which is a positive definite quantity if we choose $\varepsilon$
to be smaller than~$m$. 
We have thus shown that there exist negative frequency modes
(i.e. modes with $\omega\le -m$) which have a positive norm. 
Moreover, this occurs only when $(qV)$ is greater than~$(2m)$ 
(note that $(qV)=(2m+\varepsilon)$) which is exactly the condition 
under which Klein paradox is expected to arise.
As we had discussed in the last paragraph, it is this feature
of the Klein paradox that is responsible for exciting the
detector.

As in the case of a time-dependent electric field background,
evaluating the effective Lagrangian for an arbitrary 
time-independent electric field proves to be a difficult task
and the effective Lagrangian in such cases has been evaluated 
only for a few specific examples. 
We had pointed out above that a time-independent electric 
field is expected to produce particles only if the depth 
of the potential~$\left[qA(x)\right]$ is greater than $(2m)$.
It will be a worthwhile exercise to show that the effective 
Lagrangian has an imaginary part only under such a condition.

\subsection{In time-independent magnetic field 
backgrounds}\label{subsec:tindmf}

A time-independent magnetic field background can be described
by the vector potential
\begin{equation}
A^{\mu}=(0, 0, A(x), 0),\label{eqn:vecmag}
\end{equation}
where $A(x)$ is an arbitrary function of $x$.
This vector potential gives rise to the magnetic field 
${\bf B}= (dA/dx)\, {\hat {\bf z}}$, where ${\hat {\bf z}}$ 
is the unit vector along the positive $z$-axis.
The modes of the quantum scalar field~${\hat \Phi}$ in 
such a background can be decomposed exactly as we did in 
Eq.~(\ref{eqn:modetine}) in the case of the 
time-independent electric field background.
Hence, the transition amplitude 
${\tilde {\cal A}}_{{\bf k},{\bf l}}^*({\cal E})$ of an 
inertial detector in a time-independent magnetic field 
background will also be proportional to a delta 
function as in Eq.~(\ref{eqn:minkampnld}). 
However, on substituting the mode~(\ref{eqn:modetine}) 
and the vector potential~(\ref{eqn:vecmag}) in the scalar 
product~(\ref{eqn:sclrprdct}), we find that
\begin{equation}
(u_{\omega{\bf k}_{\perp}},u_{\omega{\bf k}_{\perp}})
=\left(2\omega\right)\, (2\pi)^2\, \delta^{(2)}(0)\,
\int\limits_{-\infty}^{\infty} dx\,
\vert f_{\omega{\bf k}_{\perp}}(x)\vert^2,
\end{equation} 
which is clearly a positive definite quantity whenever 
$\omega\ge m$.
In other words, unlike the case of the time-independent 
electric field background, in a time-independent magnetic 
field background, the definition of positive frequency modes 
{\it always}\/ matches the definition of positive norm modes.
Therefore, as in the case of an inertial detector in the
Minkowski vacuum, an inertial detector will not respond 
in the vacuum state in a time-independent magnetic field 
background.

Let us now try to evaluate the effective Lagrangian 
for an arbitrary time-independent magnetic field 
background~\cite{tunn}.
The operator ${\hat H}$ corresponding to the vector
potential~(\ref{eqn:vecmag}) is given by
\begin{equation}
{\hat H}\equiv {\partial_{t}}^2 - {\nabla}^2 
+ 2iq A(x)\, \partial_y +q^2 A^{2}(x).
\end{equation}
Using the translational invariance of the operator ${\hat H}$ 
along the time coordinate $t$ and the spatial coordinates $y$ 
and $z$, the kernel corresponding to this operator can be
written as
\begin{equation}
K({\tilde x}, {\tilde x}, s)
=\left({1 \over 4\pi s}\right)
\int\limits_{-\infty}^{\infty} {dp_y \over 2\pi}\; 
\langle x\vert e^{-i{\hat H'}s} \vert x \rangle,
\end{equation}
where
\begin{equation}
{\hat H'}\equiv -d_x^2+\left[p_y-qA(x)\right]^2.
\end{equation}
The quantity $\langle x\vert e^{-i{\hat H'}s} \vert x \rangle$ 
can now expressed using the Feynman-Kac formula as follows 
(see, for e.g., Ref.~\cite{fh65}, p.~88):
\begin{equation}
\langle x\vert e^{-i{\hat H'}s} \vert x \rangle
= \sum_{E} {\vert \Psi_E(x)\vert}^2\, e^{-iEs},\quad
{\rm where}\quad
{\hat H'}\Psi_E=E\Psi_{E},
\end{equation}  
so that $K({\tilde x}, {\tilde x}, s)$ is given by
\begin{equation}
K({\tilde x}, {\tilde x}, s)
=\left({1 \over 4\pi s}\right)
\int\limits_{-\infty}^{\infty} {d p_y \over 2 \pi}
\sum_{E} {\vert \Psi_E(x)\vert}^2\, e^{-iEs}.
\end{equation}
(It is assumed here that the summation over $E$ stands for 
integration over the relevant range when $E$ varies 
continuously.)
Since the potential term, viz. $\left[p_y-qA(x)\right]^2$, 
in the operator ${\hat H'}$ above is a positive definite 
quantity, the eigen value $E$ can only lie in the range 
$(0, \infty)$.
Substituting the above expression for 
$K({\tilde x},{\tilde x}, s)$
in Eq.~(\ref{eqn:lcorr}), we find that 
${\cal L}_{\rm corr}$ is given by
\begin{equation}
{\cal L}_{\rm corr}
= -\left(\frac{i}{4\pi}\right)
\int\limits_{-\infty}^{\infty} \frac{dp_y}{2 \pi}\,
\sum_{E} {\vert \Psi_E(x)\vert}^2\, 
\int_0^{\infty} {ds \over s^2}\, {e^{-i(m^2+E)s}}.
\end{equation}
(It should be noted here that for the case of the complex scalar 
field we are considering here, ${\cal L}_{\rm corr}$ is, in 
fact, twice the quantity defined in Eq.~(\ref{eqn:lcorr}).)
On carrying out the integral over $s$, we finally obtain 
that
\begin{equation}
{\cal L}_{\rm corr}
= \left(\frac{1}{4\pi}\right)
\int\limits_{-\infty}^{\infty}\frac{dp_y}{2\pi}\,
\sum_{E} {\vert \Psi_E(x)\vert}^2\, (m^2+E)\, 
\left[\ln (m^2+E) -1\right].
\end{equation}
Since $(m^2+E)>0$, it is easy to see from this expression 
that ${\cal L}_{\rm corr}$ is a real quantity.
Though we are unable to express the effective Lagrangian
for an arbitrary time-independent magnetic field in a closed
form, we have been able to show that it does not have an 
imaginary part which then implies that such a background 
will not produce particles.

\section{Discussion}\label{sec:dscssn}

In this concluding section, we shall first briefly summarize 
the results of the analysis we have carried out in this paper 
and then go on to discuss the implications of our analysis for 
classical gravitational backgrounds.

\subsection{What do detectors detect?}\label{subsec:what}

In order to clearly illustrate the conclusions we wish to draw 
from our analysis, we have tabulated the results we have obtained 
in the last three sections in Table~\ref{table}.

\begin{table}\label{table}
\caption{Comparison}
\begin{tabular}{cccc}
\hline
& Bogolubov & Detector & Effective \\ 
& coefficient & response & Lagrangian \\ 
& $\beta$ & ${\cal P}({\cal E})$ & 
${\rm Re.}~{\cal L}_{\rm corr} \quad
{\rm Im.}~{\cal L}_{\rm corr}$ \\
\hline
In inertial coordinates & 0 &  0 & $\!\!\!\! 0 
\qquad\qquad\;\; 0$\\
In Rindler coordinates & $\ne 0$ & $\ne 0$ 
& $\!\!\!\! 0\qquad\qquad\;\; 0$\\
In rotating coordinates & 0 & $\ne 0$ 
& $\!\!\!\! 0 \qquad\qquad\;\; 0$\\
In `cusped' coordinates & 0 & $\ne 0$ 
& $\!\!\!\! 0\qquad\qquad\;\; 0$\\
Between Casimir plates & 0 & 0 
& $\!\!\!\!\!\!\!\ne 0\qquad\quad\;\;\;\,0$\\
In a time-dependent & $\ne 0$ & $\ne 0$ 
& $\!\!\!\!\,\ne 0\qquad\quad\,\,\ne 0$\\
electric field & & &\\
In a time-independent & $\;\,\ne 0^a$
& $\ne 0$ & $\!\!\!\!\,\ne 0\qquad\quad\,\,\ne 0$\\
electric field & & &\\
In a time-independent & 0 & 0  
& $\!\!\!\!\!\!\!\ne 0\qquad\quad\;\;\;\,0$\\
magnetic field & & &\\
\hline
\end{tabular}
\tablenotes{$^a$~Actually, the Bogolubov coefficient $\beta$ is 
trivially zero in a time-independent electric field background.
We refer here to particle production that can occur in such a 
background due to Klein paradox (see Section~\ref{subsec:tindef}).} 
\end{table}

To begin with, we would like to emphasize the point we 
had discussed in detail earlier, viz. that the response 
of a detector can be non-zero even when the Bogolubov 
coefficient~$\beta$ is zero. 
The cases of the rotating detector and that of the detector 
in motion along the `cusped' trajectory clearly support this 
statement (see rows three and four, columns one and two in 
Table~\ref{table}).
Also, it is important to note that the detector response 
can be non-zero even when the effective Lagrangian vanishes 
identically---evidently, all the non-inertial cases support 
this point (cf.~rows and columns two, three and four). 
It should be stressed here that this is true even in case of 
the Rindler coordinates, a non-inertial frame in which the 
Bogolubov coefficient~$\beta$ proves to be non-zero.
Clearly, a non-zero response of a detector does not necessarily 
imply particle production.

Having said that, it is important to note that irrespective of 
its motion the response of a detector will be non-zero whenever 
there is particle production taking place.
In that sense a detector {\it is}\/ sensitive to particle production.
Moreover, if we restrict the motion of the detector to inertial 
trajectories, then we can avoid the non-inertial effects and, in 
such cases, the detector response will be non-zero {\it only}\/ 
when particle production takes place. 
The fact that an inertial detector does not respond either in
the Casimir vacuum or in a time-independent magnetic field 
(wherein the effective Lagrangian had no imaginary part, 
cf.~rows five and eight, columns two and four); whereas such a 
detector responds non-trivially both in time-dependent as well 
as time-independent electric fields (wherein the imaginary part 
of the effective Lagrangian is, in general, expected to be 
non-zero, cf.~rows six and seven, columns two and four) support 
this point.
However, as the case of the time-dependent electric field
background suggests, the response of an inertial detector 
will not necessarily be proportional to the number of 
particles produced by the background. 

\subsection{Implications for classical gravitational 
backgrounds}\label{subsec:implctns}

Unlike in flat spacetime or classical electromagnetic 
backgrounds, there exists no special frame of reference 
in a classical gravitational background and all coordinate 
systems have to be treated equivalently.  
This feature severely restricts the utility of a detector 
to study the phenomenon of particle production in a classical 
gravitational background.
Until now, we had discussed as to how the response of a
detector compares with the results from the Bogolubov 
transformations and the effective Lagrangian approach.
In what follows, we shall attempt to understand as to how 
the effective Lagrangian would behave under arbitrary
coordinate transformations.

Consider a massless and real quantum scalar field evolving 
in a gravitational background described by the metric tensor 
$g_{\mu\nu}$. 
This scalar field will satisfy an equation of motion such as 
Eq.~(\ref{eqn:mtn}) with $m$ set to zero and the operator 
${\hat H}$ given by Eq.~(\ref{eqn:gravop}).
The quantity ${\cal L}_{\rm corr}$ obtained by integrating 
out the degrees of freedom of the quantum scalar field can 
then be expressed in terms of the determinant of the operator 
${\hat H}$ (see, for e.g., Ref.~\cite{dewitt75}).
The determinant of the operator ${\hat H}$ can in turn be 
expressed as a product of its eigen values, say, $\lambda_{i}$, 
where these eigen values are obtained by solving the 
differential equation ${\hat H}w_{i}=\lambda_{i}w_{i}$ with 
respect to a complete set of modes $\{w_{i}({\tilde x})\}$.
Let us now perform a coordinate transformation on the metric
tensor $g_{\mu\nu}$.
Let the operator and its eigen values in the new coordinate 
system be ${\hat {\bar H}}$ and ${\bar \lambda}_{i}$, where 
the eigen values are now obtained by solving the eigen value 
equation 
${\hat {\bar H}}{\bar w}_{i}={\bar \lambda}_{i}{\bar w}_{i}$ 
with respect to a new set of modes $\{{\bar w}_{i}({\tilde x})\}$.
If we now assume that the new modes ${\bar w}_{i}$ are obtained 
from the old ones (viz. ${w}_{i}$) by explicitly substituting 
the corresponding coordinate transformation, then it is easy to
show that the eigen values $\lambda_{i}$ will remain unchanged 
(i.e. ${\bar \lambda}_{i}={\lambda}_{i}$). 
In such a case, the effective Lagrangian will remain invariant 
under coordinate transformations and will therefore behave as 
a scalar quantity.

Though the effective Lagrangian thus obtained will be 
invariant under coordinate transformations, it will be
a divergent quantity and we will need to regularize this 
expression.
Now, a complete set of modes can be used to evaluate the 
kernel and, the kernel in turn, can be used to obtain the 
corresponding Green function.
Therefore, choosing to work with a particular set of modes 
$\{w_{i}({\tilde x})\}$ (from which all the other sets
$\{{\bar w}_{i}({\tilde x})\}$ are obtained by explicitly 
substituting the coordinate transformations) corresponds to 
choosing a particular vacuum state for the quantum field.
In flat spacetime, divergent expressions are always regularized 
by subtracting the contribution due to the Minkowski vacuum.
So, if we choose to work with those set of modes that lead 
to the Green function in the Minkowski vacuum, then the 
regularized effective Lagrangian will be trivially zero in 
all coordinates in flat spacetime. 
In fact, this is exactly what we have found from our analysis. 
We found that the kernel that leads to the Green function in 
the Minkowski vacuum is invariant under coordinate transformations 
in the coincident limit (cf.~Eqs.~(\ref{eqn:minkkrnl}), 
(\ref{eqn:rindkrnl}), (\ref{eqn:rotkrnl}), 
(\ref{eqn:cuspkrnl})) and, hence, the 
corresponding effective Lagrangian identically reduced to
zero in all the non-inertial coordinates on regularization.
 
However, these arguments do not still imply that the effective
Lagrangian will be unique in a given gravitational background.
Instead of choosing to work with modes that led to the Green 
function in the Minkowski vacuum, we could have chosen to work 
with modes that lead to the Green function in the Rindler vacuum. 
If we now use these modes to evaluate the kernel, then the 
effective Lagrangian corresponding to this kernel will be 
different from the effective Lagrangian that corresponds 
to the Minkowski vacuum and, hence, will lead to a non-zero 
value on regularization. 
In fact, when the contribution due to the Minkowski vacuum is 
subtracted from the effective Lagrangian in the Rindler vacuum, 
one obtains a non-zero and real quantity which has a thermal 
nature (see Ref.~\cite{cr76}; also see Ref.~\cite{paddyup} in 
this context).

This result can be stated in a more formal and general terms 
along the following lines, which will prove to be useful. 
We notice that the essential physics of a free field theory 
is contained in the two-point function $G({\tilde x}, 
{\tilde x'})=\langle 0\vert {\rm T}\left[\Phi({\tilde x})\, 
\Phi({\tilde x'})\right] \vert 0 \rangle$ (where ${\rm T}$ 
denotes time-ordering), which satisfies an inhomogeneous 
differential equation. 
But the same differential equation will be satisfied by a 
function ${\cal F}({\tilde x},{\tilde x'})=\langle {\bar \Psi} 
\vert {\rm T}\left[\Phi({\tilde x})\, \Phi({\tilde x'})\right]
\vert {\bar \Psi}\rangle$ defined with respect to any normalizable 
quantum state $\vert{\bar \Psi}\rangle$. 
In particular, if there exist two different vacuum states, then 
the corresponding two-point functions will differ, i.e. they 
will not be related by a coordinate relabelling appropriate for 
a biscalar (which is precisely what happens in the case of 
Rindler and Minkowski vacuum states). 
But, since the functions $G({\tilde x}, {\tilde x'})$ and
${\cal F}({\tilde x},{\tilde x'})$ satisfy the same 
inhomogeneous differential equation, they will, in general, 
differ by a solution to the homogeneous equation. 
Alternatively, they will differ by the boundary conditions 
both at the asymptotic regimes as well as near horizons that 
the spacetime may contain. 
(One popular way of choosing the boundary condition in standard 
quantum field theory is through Euclidean continuation which, 
of course, will not work in a general curved spacetime.)

The above discussion highlights the key difficulty:~unless 
external criteria are imposed to choose a particular boundary 
condition, the class of functions ${\cal F}({\tilde x}, 
{\tilde x'})$ are valid two-point functions of the theory, a 
priori. 
In order to choose one (or a few) of them as special, we need to 
study their general behavior and impose some boundary conditions. 
In fact, not all of them will lead to quantum field theories 
which are unitarily equivalent.  
It has been shown in literature that there exists no unitary 
transformation relating the Fock space constructed from the 
Minkowski vacuum and the Fock space determined by the Rindler 
vacuum~\cite{gerlach89}.
(Evidently, it is this feature that leads to the inequivalent 
quantization and, as a result, the non-zero effective Lagrangian 
in the Rindler vacuum.)
This result points to the fact that in an arbitrary gravitational 
background not all coordinate transformations can be implemented 
unitarily.
This implies that, in general, there exist families of 
inequivalent Fock spaces in a curved spacetime (see, for 
e.g., Ref.~\cite{fulling77}).
In the case of flat spacetime, the Fock space associated with 
the Minkowski vacuum provides us with a natural basis.
But, no such special Fock space seems to be available to us 
in a curved spacetime.
In such a situation, which of the inequivalent Fock spaces 
should we choose to work with?
Will we be able to choose one of these Fock spaces on our own 
or will it be chosen automatically when we set up an experiment? 

An important aspect of the modes associated with the Minkowski
coordinates in flat spacetime are that they are well defined 
and regular in the entire region of the spacetime. 
Recently, it has been argued that this feature should be utilized 
to propose a possible criterion for selecting a particular vacuum 
state (and the associated Fock space) from amongst the different 
possibilities in a curved spacetime~\cite{whrp99}. 
The requirement that the modes be regular {\it throughout}\/ the 
spacetime (so that the states can evolve from data in the infinite 
past) has been suggested as the physical criterion to distinguish
between the different states.
(This criterion immediately picks out the Minkowski vacuum 
state in flat spacetime as other states such as the Rindler 
vacuum lead to divergences on the horizons.)
 
It is, however, not clear whether this condition may turn out
to be overly restrictive. 
In spacetimes with horizons, two-point functions for different 
vacua will have different---and sometimes singular---behavior 
at the horizon. 
For example, the Minkowski coordinates of flat spacetime is 
similar to the Kruskal coordinates of Schwarzschild spacetime; 
the analogue of the Minkowski vacuum in the Schwarzschild 
spacetime will be the Hartle-Hawking vacuum. 
We, however, have physical situations which are best described 
with respect to the Unruh vacuum (corresponding to a collapsing 
star) or even the Boulware vacuum (around a static star) in the 
Schwarzschild spacetime. 
It is probably better to classify the boundary conditions and 
try to identify a class of vacuum states rather than impose 
regularity throughout the spacetime.

A closely related issue is the distinction between a change of 
reference frame and coordinate relabelling. 
If one deals with tensorial quantities, coordinate relabelling 
does not lead to any new physical effects. 
One can certainly use the Minkowski modes in the Rindler frame 
(after expressing the Minkowski coordinates in the modes in terms 
of the Rindler coordinates) to define the Minkowski vacuum state 
and carry out quantum field theory in the  Rindler frame.
The results will be completely equivalent to field theory in the 
Minkowski coordinates---albeit expressed in a strange language. 
When one uses the terminology ``change of reference frame" one 
has something different in mind---though its exact definition 
is difficult to express in general terms. 
In simple contexts like the Minkowski and Rindler coordinates, 
one implies changing over the description to a language which 
is {\it natural}\/ to the coordinates that have been chosen 
(such as, for e.g., choosing to work with positive norm modes 
defined with respect to the new time coordinate). 
There is an important distinction which arises between the 
electromagnetic and gravitational fields in this context, 
which we shall now briefly describe.

Let us consider a laboratory experiment in which a pair of 
parallel capacitor plates are set up with a given potential 
difference between them, corresponding to an electric field. 
If the field is strong enough, we should see pair production 
between the plates. 
The pairs produced will move towards the plates and will try 
to reduce the charge densities on the plates thereby reducing 
the strength of the electric field between them. 
To maintain the original strength of the electric field, the 
external source has to do work which will supply the energy 
of the particles produced by the electric field. 
Note that, nowhere in this description did we need to specify 
the gauge used to describe the electric field, even though to 
set up the quantum field theory and obtain the pair production 
rate in the electric field, one might choose to work in a 
specific gauge. 
The key reason for this result---which is not often emphasized---is 
that the source of electromagnetic field, viz. the electric current 
$J^{\mu}$, is {\it gauge-invariant rather than merely being 
gauge-covariant}; i.e. $J^{\mu}$ is a ``scalar'' under gauge 
transformations (unlike, for example, the charged scalar field, 
which is only gauge-covariant and picks up a phase factor under 
a gauge transformation). 
We can therefore specify the experimental set up in terms of 
charges and currents and ask what will happen in the laboratory 
without ever concerning ourselves about the gauge. 

The situation is quite different in the case of gravity. 
The analogue of the gauge transformation in gravity is the
coordinate transformation. 
But, the source of the classical gravitational field, viz. 
the stress-energy tensor $T^{\alpha\beta}$, is only a covariant 
quantity rather than an invariant one. 
This feature, of course, makes no difference in classical 
theory. 
We may choose any coordinate system we like to specify the 
components of the stress-energy tensor and solve the Einstein's 
equations to obtain the metric; if we change the coordinates, 
then, both the stress-energy tensor and the metric will change 
suitably, maintaining general covariance at the classical level.
The situation is different in quantum theory where the
vacuum state, for instance, can be different based on 
the modes which are chosen. 
Since, different sets of modes may be natural for different 
coordinate frames---corresponding to different metric and 
stress-energy tensor components---we cannot phrase questions 
in the case of gravity in a manner similar to the 
electromagnetic case (unless one could reformulate Einstein's 
equations entirely in terms of scalar invariants which seems 
to be an impossible task).

Thus, we come to the inevitable conclusion that an extra 
prescription---say, in terms of the boundary conditions on the 
two-point function---is required in an arbitrary spacetime to 
define and deal with issues such as particle production. 
We conjecture that in spacetimes without horizons, this could 
be achieved with asymptotic boundary conditions, whereas in 
spacetimes with horizons, we may also need to specify the 
behavior of the modes on the horizon as well. 
In fact, it should be possible to arrive at some general 
conclusions regarding the behavior of two-point functions 
in arbitrary spacetimes along these lines.
We hope to address these issues further in a future publication.

\appendix{Evaluating the Feynman propagator}\label{app:fgfn}

In this appendix, we shall evaluate the Feynman propagator for 
the case of a massless scalar field in the three non-inertial 
coordinate systems we had discussed in Section~\ref{sec:nifrms}. 

Before we go on to evaluate the Feynman propagator in the 
non-inertial coordinate systems, let us first consider 
the case of the Minkowski coordinates. 
In these coordinates the operator ${\hat H}$ as defined
in Eq.~(\ref{eqn:gravop}) is given by
\begin{equation}
{\hat H}\equiv\left(\partial_{t}^2- {\bf {\nabla}}^2\right).
\end{equation}
This is the time evolution operator of a free quantum mechanical
particle and the kernel~(\ref{eqn:kernel}) corresponding to such 
an operator is given by (see, for e.g., Ref.~\cite{fh65}, p.~42)
\begin{equation}
K({\tilde x},{\tilde x'};s)
=\left(\frac{1}{16 \pi^2 i s^2}\right)\;
\exp-\left(\frac{i}{4s}\right)
\left[(t-t')^2-\vert {\bf x}-{\bf x}'\vert^2\right].
\label{eqn:minkkrnl}
\end{equation}
On substituting this kernel in Eq.~(\ref{eqn:fgfn}) and
evaluating the resulting integral, we find that the Feynman 
propagator in the Minkowski coordinates is given by
\begin{eqnarray}
G_{\rm F}({\tilde x},{\tilde x'})
=\left(\frac{i}{4\pi^2}\right)
\left(\frac{1}{\left[(t-t')^2-i\epsilon\right]
-\vert {\bf x}-{\bf x}'\vert^2}\right)
\label{eqn:fgfnine}
\end{eqnarray}
which, apart from a factor of $i$, is the same as the 
Wightman function (\ref{eqn:wgfnine}) {\it provided}\/  
we modify the quantity $\left[(t-t')^2-i\epsilon\right]$ 
to $(t-t'-i\epsilon)^2$. 
(The factor~$i$ arises because the Feynman propagator 
is~$(-i)$ times the vacuum expectation value of the 
time-ordered product of the quantum field.)

\newpage
\section{In the Rindler frame}\label{app:rind}

The operator~${\hat H}$ (as defined in~Eq.~(\ref{eqn:gravop})) 
corresponding to the Rindler metric~(\ref{eqn:rindmet}) is 
given by
\begin{equation}
{\hat H}\equiv\left(\frac{1}{\xi^2}\partial_{\eta}^2
-\frac{g^2}{\xi}\partial_{\xi}\left(\xi\,\partial_{\xi}\right)
-\partial_y^2-\partial_z^2\right).
\end{equation}
This operator is invariant under translations along the~$y$ 
and the $z$ directions. 
In other words, along these two directions, the kernel 
corresponds to that of a free particle.
Exploiting this feature, we can write the quantum mechanical
kernel corresponding to the case $y=y'$ and $z=z'$ as
\begin{equation}
K({\tilde x}, {\tilde x'} ;s)
=\left(\frac{1}{4\pi is}\right)\;
\langle \eta, \xi \vert  
e^{-i{\hat H'}s}\vert \eta', \xi'\rangle,
\end{equation}
where the operator~${\hat H'}$ is given by
\begin{equation}
{\hat H'}\equiv \left(\frac{1}{\xi^2}\partial_{\eta}^2
-\frac{g^2}{\xi}\partial_{\xi}
\left(\xi\,\partial_{\xi}\right)\right).
\end{equation}
On rotating the time coordinate $\eta$ to the negative 
imaginary axis (i.e. on setting $\eta=-i\eta_E$) and 
changing variables to $u=\left(g^{-1}\xi\right)$, we 
find that
\begin{equation}
{\hat H'}\equiv
\left(-\frac{1}{g^2u^2}\partial_{{\eta}_E}^2
-\frac{1}{u}\partial_{u}\left(u\, \partial_{u}\right)\right).
\end{equation}
If we now identify $u$ as a radial variable and $(g\eta_E)$ 
as an angular variable, then the operator~${\hat H'}$ is
similar in form to the Hamiltonian operator of a free particle 
in polar coordinates (in 2-dimensions)~\cite{tv77,cd78,tv79}. 
The kernel corresponding to this operator can then be written as
\begin{equation}
\langle \eta, \xi \vert  
e^{-i{\hat H'}s}\vert \eta', \xi'\rangle
=\left(\frac{1}{4\pi s}\right)\;
\exp \left(\frac{i}{4 g^{2} s}\right)
\left(\xi^2+\xi'^2 -2\xi\xi' {\rm cosh}
\left[g(\eta-\eta')\right]\right).
\end{equation} 
Therefore, when $\xi=\xi'$, the complete kernel is given by
\begin{equation}
K({\tilde x}, {\tilde x'} ;s)
=\left(\frac{1}{16\pi^2 is^2}\right)\;
\exp -\left(\frac{i \xi^2}{g^2 s}\right)
\left({\rm sinh}^2\left[g(\eta-\eta')/2\right]\right)
\label{eqn:rindkrnl}
\end{equation}
and the Feynman propagator corresponding to this kernel can be 
easily evaluated to be
\begin{eqnarray}
\!\!\!\!\!\!
G_{\rm F}({\tilde x}, {\tilde x'})
&=& \left(\frac{i g^2}{16\pi^2 \xi^2}\right)\;
\left({\rm sinh}^{-2}\left[g(\eta-\eta')/2\right]
+i\epsilon\right)\nonumber\\
&=& \left(\frac{i}{4\pi^2\xi^2}\right)
\sum_{n=-\infty}^{n=\infty} 
\left[\left(\eta-\eta'+2\pi i n g^{-1}\right)^{2}
-i\epsilon\right]^{-1}.
\label{eqn:fgfnrind}
\end{eqnarray}

\section{In the rotating coordinates}\label{app:rot}

The operator ${\hat H}$ corresponding to the 
metric~(\ref{eqn:rotmet}) in the rotating coordinates 
is given by
\begin{equation}
{\hat H}\equiv
\left(\left(\partial_{t}-\Omega\, \partial_{\theta}\right)^2 
-\frac{1}{r}\partial_{r}\left(r\, \partial_{r}\right)
-\frac{1}{r^2}\partial_{\theta}^2-\partial_{z}^2\right).
\end{equation}
Exploiting the translational invariance of this operator along
the $t$, $z$ and the $\theta$ directions, we can write the
kernel corresponding to this operator for the case $r=r'$, 
$\theta=\theta'$ and $z=z'$ as follows:
\begin{eqnarray}
\!\!\!\!\!\!\!\!\!\!\!\!\!\!\!\!\!
K({\tilde x}, {\tilde x'};s)
&=&\left(\frac{1}{\sqrt{4\pi is}}\right)
\left(\frac{1}{2\pi}\right)\nonumber\\ 
& &\times\;\sum_{m=-\infty}^{\infty}\;
\int\limits_{-\infty}^{\infty}\frac{d\omega}{2\pi}\,
e^{-i\omega(t-t')}\; e^{i(\omega+m\Omega)^2 s}\; 
\langle r \vert e^{-i{\hat H'}s}\vert r \rangle,
\end{eqnarray}
where~${\hat H'}$ is given by 
\begin{equation}
{\hat H'}\equiv \left(-d_{r}^2-\frac{1}{r}d_{r} 
+\frac{m^2}{r^2}\right).
\end{equation}
On carrying out the integral over~$\omega$, we obtain that
\begin{eqnarray}
\!\!\!\!\!\!\!\!\!\!\!\!
K({\tilde x},{\tilde x'};s)
&=&\left(\frac{1}{8\pi^2 s}\right)\;
e^{-\left[i(t-t')^2/4s\right]}\nonumber\\
& &\qquad\qquad\qquad\times\;
\sum_{m=-\infty}^{\infty}\; 
e^{im\Omega(t-t')}\;
\langle r \vert e^{-i{\hat H'}s}\vert r\rangle.
\label{eqn:rotkerpen}
\end{eqnarray}
The normalized modes of the operator~${\hat H'}$ 
corresponding to an energy eigen value $E=q^2$ 
are given by (cf.~Ref.~\cite{arfken85}, p.~591)
\begin{equation}
\Psi_{q}(r)=\sqrt{q}\; J_{m}(qr),
\end{equation}
where $J_{m}$ is a Bessel function of integral order 
and $q$ runs continuously from zero to~$\infty$.
The kernel~$\langle r \vert e^{-i{\hat H'}s}\vert r\rangle$ 
can now be expressed in terms of these modes using the 
Feynman-Kac formula as follows (see, for instance, 
Ref.~\cite{fh65}, p.~88):
\begin{equation}
\langle r \vert e^{-i{\hat H'}s}\vert r \rangle
=\int\limits_{0}^{\infty}dq\, q\, J_{m}^2(qr)\, e^{-iq^2s}.
\end{equation}
On substituting this expression in Eq.~(\ref{eqn:rotkerpen}), 
we obtain that
\begin{eqnarray}
& &\!\!\!\!\!\!\!\!\!\!\!\!\!\!\!\! 
\lefteqn{K({\tilde x},{\tilde x'};s)}\nonumber\\
& &=\;\left(\frac{1}{8\pi^2 s}\right)
e^{-\left[i(t-t')^2/4s\right]}
\sum_{m=-\infty}^{\infty} e^{im\Omega(t-t')}\;
\int\limits_{0}^{\infty}dq\, q\, J_{m}^2(qr)\, 
e^{-iq^2s}\nonumber\\
& &=\;\left(\frac{1}{8\pi^2 s}\right)
e^{-\left[i(t-t')^2/4s\right]}\nonumber\\
& &\qquad\;\;\times\;
\biggl\{\int\limits_{0}^{\infty}dq\, q\, J_{0}^2(qr)\, 
e^{-iq^2s}\nonumber\\
& &\qquad\qquad\quad
+\; 2\sum_{m=1}^{\infty}\cos\left[m\Omega(t-t')\right]\,
\int\limits_{0}^{\infty}dq\, q\, J_{m}^2(qr)\, 
e^{-iq^2s}\biggl\}.
\end{eqnarray}
The integrals over $q$ can be expressed in terms of 
modified Bessel functions~$I_{m}$ as follows (see, 
for e.g., Ref.~\cite{pbm86}, p.~223):
\begin{eqnarray}
& &\!\!\!\!\!\!\!\!\!\!\!\!\!\!\!\!\!\!\!\!
\lefteqn{K({\tilde x},{\tilde x'};s)}\nonumber\\
&=&\left(\frac{1}{16 \pi^2 i s^2}\right)
\exp-\left(\frac{i}{4s}\right)
\left[(t-t')^2-2 r^2\right]\nonumber\\
& &\;\;\times\;
\left\{I_{0}(r^2/2is)
+2\sum_{m=1}^\infty 
\cos\left[m\Omega(t-t')\right]\, I_{m}(r^2/2is)\right\}.
\end{eqnarray}
On using the identity (cf.~Ref.~\cite{pbm86}, p.~695) 
\begin{equation}
\sum_{k=1}^{\infty}\cos(ka)\, I_{k}(z)
=\frac{1}{2}\left[e^{z\cos(a)} -I_{0}(z)\right],
\end{equation}
we finally obtain that
\begin{equation}
K({\tilde x}, {\tilde x'};s)
=\left(\frac{1}{16 \pi^2 i s^2}\right)\;
\exp -\left(\frac{i}{4s}\right)\left[(t-t')^2
-4r^2 \sin^2\left[\Omega(t-t')/2\right]\right].
\label{eqn:rotkrnl}
\end{equation}
The corresponding Feynman propagator is given by 
\begin{equation}
G_{\rm F}({\tilde x}, {\tilde x'})
=\left(\frac{i}{4\pi^2}\right)
\left(\frac{1}{(t- t')^2
-4r^2 \sin^2\left[\Omega(t-t')/2\right]- i \epsilon}\right).
\label{eqn:fgfnrot}
\end{equation}

\section{In the `cusped' coordinates}\label{app:cusp} 

The operator~${\hat H}$ corresponding to the line
element~(\ref{eqn:cuspmet}) is given by
\begin{equation}
{\hat H}\equiv 
\left(2\, \partial_{\bar t}\, 
\partial_{\bar y}-\partial_{\bar x}^2
-2a\, {\bar x}\, \partial_{\bar y}^2
-\partial_{z}^2\right).
\end{equation}
Exploiting the translational invariance along the 
${\bar t}$, ${\bar y}$ and ${\bar z}$ directions, 
we can write the kernel corresponding to this 
operator for the case ${\bar x}={\bar x'}$,
${\bar y}={\bar y'}$ and $z=z'$ as
\begin{equation}
K({\tilde x},{\tilde x'};s)
=\left(\frac{1}{\sqrt{4\pi is}}\right) 
\int\limits_{-\infty}^{\infty}\frac{d\omega}{2\pi}\; 
e^{-i\omega({\bar t}-{\bar t'})}\; 
\int\limits_{-\infty}^{\infty}\frac{dp_{\bar y}}{2\pi}\;
e^{-2i\omega p_{\bar y} s}\;
\langle {\bar x}\vert e^{-i{\hat H'}s}
\vert {\bar x}\rangle,
\end{equation}
where the operator ${\hat H'}$ is given by
\begin{equation}
{\hat H'}\equiv
-d_{\bar x}^2 + 2\, a\, p_{\bar y}^2\, {\bar x}. 
\end{equation}
On integrating over $\omega$, we obtain that
\begin{equation}
\!\!\!\!\!\!\!\!\!\!
\!\!\!\!\!\!\!
K({\tilde x},{\tilde x'};s)
=\left(\frac{1}{\sqrt{4\pi is}}\right)\;
\int\limits_{-\infty}^{\infty}\frac{dp_{\bar y}}{2\pi}\;
\delta^{(1)}\left[2p_{\bar y} s-({\bar t}-{\bar t'})\right]\;\,
\langle {\bar x}\vert 
e^{-i{\hat H'} s}\vert {\bar x}\rangle.
\label{eqn:cuspkrnlpen}
\end{equation}
The operator ${\hat H'}$ above corresponds to that of a particle
in a linear potential. 
The kernel corresponding to this case is well-known and is 
given by (see, for instance, Ref.~\cite{dm94}, p.~194)
\begin{equation}
\langle {\bar x}\vert e^{-i{\hat H'}s}\vert {\bar x}\rangle
=\left(\frac{1}{\sqrt{4\pi is}}\right)
\exp -\left(\frac{i}{3}\right)\left(6a p_{\bar y}^2{\bar x}s
+a^{2} p_{\bar y}^4 s^3\right).
\end{equation}
On substituting this expression in Eq.~(\ref{eqn:cuspkrnlpen}), 
we find that the complete kernel is given by
\newpage
\begin{eqnarray}
& &\!\!\!\!\!\!
\lefteqn{K({\tilde x},{\tilde x'};s)}\nonumber\\
& &\,=\;\left(\frac{1}{4\pi is}\right)
\int\limits_{-\infty}^{\infty}\frac{dp_{\bar y}}{2\pi}\;
\delta^{(1)}\left[2p_{\bar y} s-({\bar t}-{\bar t'})\right]\;
\exp -\left(\frac{i}{3}\right)
\left(6a p_{\bar y}^2{\bar x}s
+a^{2} p_{\bar y}^4 s^3\right)\nonumber\\
& &\,=\;\left(\frac{1}{16\pi^2 is^2}\right)\;
\exp-\left(\frac{i}{48 s}\right)\left[24 a {\bar x}
({\bar t}-{\bar t'})^2 +a^2({\bar t}-{\bar t'})^4\right].
\label{eqn:cuspkrnl}
\end{eqnarray}
The resulting Feynman propagator can then be easily 
evaluated to be 
\begin{equation}
G_{\rm F}({\tilde x}, {\tilde x'})
=\left(\frac{3i}{\pi^2}\right)
\left(\frac{1}{24 a{\bar x}({\bar t}-{\bar t'})^2
+a^2({\bar t}-{\bar t'})^4 - i \epsilon}\right).
\label{eqn:fgfncusp}
\end{equation}

\begin{acknowledgments}

L.S. would like to thank Profs.~Jacob Bekenstein, Don Page and 
Valeri Frolov for discussions. 
L.S. was supported by the Israel Science Foundation (established 
by the Israel Academy of Sciences) and by the Natural Sciences 
and Engineering Research Council of Canada.

\end{acknowledgments}

\end{article}
\end{document}